\documentclass[fleqn,12pt,twoside]{article}
\usepackage[headings]{espcrc1}

\readRCS
$Id: espcrc1.tex,v 1.2 2004/02/24 11:22:11 spepping Exp $
\usepackage[dvips]{graphics,graphicx,epsfig}
\usepackage[figuresright]{rotating}
\usepackage{fancybox}
\usepackage{latexsym}
\usepackage{amssymb}
\usepackage[all]{xy}
\usepackage{color}

\newtheorem{proposition}{Proposition}

\def\Nset{\mathrm{I\!N}}
\def\Rset{\mathrm{I\!R}}

\def\rien{\rule{0pt}{0pt}}
\def\text#1{\mbox{#1}}
\def\counterfact{>}
\def\gonextline{\\\rien\hfill}
\def\seqset{\mathbf{Seq}}

\def\seqcalcZ#1{\begin{array}{@{}c@{}}\rien\vspace{-15pt}\\\displaystyle\frac{}{#1}\vspace{-17pt}\\\rien\end{array}}
\def\seqcalc#1#2{\begin{array}{@{}c@{}}\rien\vspace{-15pt}\\\displaystyle\frac{#1}{#2}\vspace{-17pt}\\\rien\end{array}}
\def\seqcalcT#1#2#3{\begin{array}{@{}c@{}}\rien\vspace{-15pt}\\\displaystyle\frac{#1\qquad#2}{#3}\vspace{-17pt}\\\rien
\end{array}}
\newcommand{\AmS}{{\protect\the\textfont2
  A\kern-.1667em\lower.5ex\hbox{M}\kern-.125emS}}

\title{Deterministic Bayesian Logic}

\author{Fr\'ed\'eric Dambreville\address{D\'el\'egation G\'en\'erale pour l'Armement, DGA/CEP/GIP\\
16 Bis, Avenue Prieur de la C\^ote d'Or\\
F 94114, France\\[3pt]
{\tt submit@FredericDambreville.com}\\
{\tt http://www.FredericDambreville.com}}%
}      

\runtitle{Deterministic Bayesian Logic}
\runauthor{F. Dambreville}

\begin{document}

\newcommand{\keywordsname}{Keywords}
\newenvironment{keywords}%
  {\small
    \list{}{\labelwidth0pt
      \leftmargin0pt \rightmargin\leftmargin
      \listparindent\parindent \itemindent0pt
      \parsep0pt
      \let\fullwidthdisplay\relax}%
    \item[\hskip\labelsep\bfseries\keywordsname:]}{\endlist}


\maketitle
\begin{center}
\rien\\
\rule{130pt}{1pt}\hfill\rien\\
$\phi\times\phi\vdash\phi,\neg\phi$\hfill\rien
\\
\emph{Freed from myself, I am the \emph{all} or the \emph{none}}\hfill\rien
\vspace{-4pt}\\
\rule{70pt}{1pt}\hfill\rien\\\rien
\end{center}
%
\begin{abstract}
In this paper a conditional logic is defined and studied.
This conditional logic, Deterministic Bayesian Logic, is constructed as a deterministic counterpart to the (probabilistic) Bayesian conditional.
The logic is unrestricted, so that any logical operations are allowed.
This logic is shown to be non-trivial and is not reduced to classical propositions.
The Bayesian conditional of DBL implies a definition of logical independence.
Interesting results are derived about the interactions between the logical independence and the proofs.
A model is constructed for the logic.
Completeness results are proved.
It is shown that any unconditioned probability can be extended to the whole logic DBL.
The Bayesian conditional is then recovered from the probabilistic DBL.
At last, it is shown why DBL is compliant with Lewis triviality.
\end{abstract}

%
%
\section{Introduction}
Bayesian inference is a powerful principle for modeling and manipulating probabilistic informations.
In many cases, Bayesian inference is considered as an optimal and legitimate rule for inferring such informations.
\begin{itemize}
\item Bayesian filters for example are typically regarded as optimal filters~\cite{arulampalam02tutorial},
\item Bayesian networks are particularly powerful tools for modeling uncertain informations.
By merging independence priors to the logical priors, Bayesian Networks are generally associated to Markovian properties, which allow quite efficient computations~\cite{pearlRus,murphy,Dambreville:cepomdp}.
\end{itemize}
Although Bayesian inference is an established principle, it is recalled~\cite{jaynes} that it has been disputed until the middle of the XXth century, in particular by the frequencist community.
What made the Bayesian inference established is chiefly a logical justification of the rule~\cite{cox,jaynes}.
Some convergence with the frequencist interpretation achieved this acceptation.
Cox derived the characteristics of Probability and of Bayesian conditional from hypothesized axioms about the probabilistic system, which themselves were reduced in terms of functional equations.
Typical axioms are:
\begin{itemize}
\item The operation which maps the probability of a proposition to the probability of its negation is idempotent,
\item The probability of $A\wedge B$ depends only of the probability of $A$ and of the probability of $B$ given that $A$ is true,
\item The probability of a proposition is independent of the way it is deduced (consistency).
\end{itemize}
It is noticed that Cox interpretation has been criticized recently for some imprecision and reconsidered~\cite{debrucq2,Halpern}.
\\[5pt]
In some sense, Cox justification of the Bayesian conditional is not entirely satisfactory, since it is implicit: it justifies the Bayesian conditional as the operator fulfilling some natural properties, but does not construct a full underlying logic  priorly to the probability.
The purpose of this paper is to construct an explicit logic for the Bayesian conditional as a conditional logic:
\begin{enumerate}
\item Build a (deterministic) conditional logic, \emph{priorly to any notion of probability}.
This logic will extend the classical propositional logic (unconditioned propositions).
It will contain conditional propositions $(\psi|\phi)$ built for any propositions $\phi$ and $\psi$,
\item Being given a probability $p$ over the unconditioned propositions, derive the probabilistic Bayesian conditional from an extension $\overline{p}$ of $p$ over the whole logic.
The Bayesian conditional will be recovered by setting $p(\psi|\phi)=\overline{p}\bigl((\psi|\phi)\bigr)$\,.
\end{enumerate}
The construction of an explicit underlying logic provides a better understanding of the Bayesian conditional, but will also make possible the comparison with other rules for manipulating probabilistic informations, based on other logics~\cite{dambreville}.
\\[5pt]
It is known that the construction of such underlying logic is heavily constrained by Lewis triviality \cite{lewis,hajek1,fraassen}, which has shown some critical issues related to the notion of conditional probability.
In particular, Lewis result implies strong hypotheses about the nature of the conditionals.
Essentially, the conditionals have to be constructed outside the space of unconditioned propositions. 
This result implied the way the logic of Bayesian conditional has been investigated.
Many approaches does not distinguish the Bayesian conditional from probabilistic notions.
This is the case of the theory called \emph{Bayesian Logic}~\cite{BayesianLogic}, which is an extension of probabilistic logic programming by the way of Bayesian conditioning.
Other approaches like conditional algebra or conditional logic result in the construction of conditional operators, which finally arise as abstraction independent of any probability.
However, these logical constructions are still approximations of the Bayesian conditional or are constrained in use. 
\\[5pt]
Since Lewis triviality is a fundamental reference in this work, it is introduced now.
By the way, different logical approaches of the Bayesian conditional are mentioned, and it is shown how these approaches avoid the triviality.
\paragraph{Lewis triviality.}
Let $\Omega$ be the set of all events, and $\mathcal{M}$ be the set of measurable subsets of $\Omega$.
Let $\mathit{Pr}(\mathcal{M})$ be the set of probability measures on $\mathcal{M}$.
Lewis triviality may be expressed as follows:
\\[5pt]
\emph{Let $A,B\in \mathcal{M}$ with $\emptyset\subsetneq B\subsetneq 
A\subsetneq \Omega$\,.
Then, it is impossible to build a proposition $(B|A)\in \mathcal{M}$ such that $\pi\bigl((B|A)\bigr)=\pi(B|A)\stackrel{\Delta}{=}\frac{\pi(A\cap B)}{\pi(A)}$ for any $\pi\in\mathit{Pr}(\mathcal{M})$\,.}
\\[5pt]
Lewis triviality thus makes impossible the construction of a (Bayesian) conditional operator within the same Boolean space.
\begin{description}
\item[Proof.]
Let $\pi$ be a probability such that $0<\pi(B)<\pi(A)<1$\,; \emph{the existence of $\pi$ is ensured by hypothesis $\emptyset\subsetneq B\subsetneq 
A\subsetneq \Omega$}\,.
\\[5pt]
For any propositions $C,D\in\mathcal{M}$\,, define $\pi_C(D)=\pi(D|C)=\frac{\pi(C\cap D)}{\pi(C)}$\,, when $\pi(C)>0$.\\
Lewis' triviality relies on the following calculus, derived when $\pi(A\cap C)>0$\,:
\begin{equation}
\label{Eq:DBL:v2:Lewis:1}
\begin{array}{@{}l@{}}\displaystyle\vspace{5pt}
\pi((B|A)|C)=\pi_C((B|A))=\pi_C(B|A)=\frac{\pi_C(A\cap B)}{\pi_C(A)}
\\\displaystyle
\rien\hspace{100pt}=\frac{\frac{\pi(C\cap A\cap B)}{\pi(C)}}{\frac{\pi(A\cap C)}{\pi(C)}}= \frac{\pi(C\cap A\cap B)}{\pi(A\cap C)}=\pi(B|C\cap A)\;.
\end{array}
\end{equation}
Denote $\sim B=\Omega\setminus B$.\\
$B\subset A$ and $0<\pi(B)<\pi(A)$ imply $\pi(A\cap B)>0$ and $\pi(A\cap \sim B)>0$\,.
\\
Then, it is inferred:
$$
\begin{array}{@{}l@{}}\displaystyle
\frac{\pi(B)}{\pi(A)}=\frac{\pi(A\cap B)}{\pi(A)}=\pi(B|A)=\pi((B|A))=\pi((B|A)|B)\pi(B)+\pi((B|A)|\sim B)\pi(\sim B)
\vspace{3pt}\\\displaystyle
\hspace{15pt}=\pi(B|B\cap A)\pi(B)+\pi(B|\sim B\cap A)\pi(\sim B)=1\times \pi(B)+0\times \pi(\sim B)=\pi(B)\;,
\end{array}
$$
which contradicts the hypotheses $0<\pi(B)$ and $\pi(A)<1$\,.
\item[$\Box\Box\Box$]\rien
\end{description}
In fact, the derivation~(\ref{Eq:DBL:v2:Lewis:1}) relies on the hypothesis that $\pi((B|A)|C)$ is defined as $\pi_C((B|A))$.
This hypothesis is necessary when $(B|A)\in\mathcal{M}$\,, but could be avoided when $(B|A)\not\in\mathcal{M}$\,.
\\[5pt]
More precisely, while the proposition $(B|A)$ is outside $\mathcal{M}$, it becomes necessary to build for any probability $\pi$ its extension $\overline{\pi}$ over the outside propositions; in particular, it will be defined $\overline{\pi}\bigl((B|A)\bigr)=\pi(B\cap A)/\pi(A)$ for any $A,B\in\mathcal{M}$\,.
But there is no reason to have $\overline{\pi_C}(D)=\overline{\pi}\bigl((D|C)\bigr)$ for $D\not\in\mathcal{M}$\,. Thus, the above triviality does not work necessarily.
\\[5pt]
The property $\overline{\pi_C}\ne\overline{\pi}\bigl((\cdot|C)\bigr)$ is somewhat counter-intuitive.
In particular, it means that conditionals are not conserved by conditional probabilities.
However, it allows the construction of a conditional logic for the Bayesian conditional; our work provides an example of such construction.
\paragraph{Probabilistic logic and Bayesian logic.}
%
Probabilistic logic, as defined by Nilsson~\cite{nilss}, has been widely studied in order to model and manipulate the uncertain information.
It tracks back from the seminal work of Boole~\cite{boole}.
In probabilistic logic, the knowledge, while logically encoded by classical propositions, is expressed by means of constraints on the probability over these propositions.
For example, the knowledge over the propositions $A,B$ may be described by:
\begin{equation}
\label{Eq:DBL:v2:BL:1}
v_1\le p(A)\le v_2\quad\mbox{and}\quad v_3\le p(A\rightarrow B)\le v_4 \;,
\end{equation}
where $v_i|1\le i\le 4$ are known bound over the probabilities.
Equations like (\ref{Eq:DBL:v2:BL:1}) turn out to be a linear set of constraints over $p$, while considering the generating propositions $A\wedge B,A\wedge \neg B,\neg A\wedge B,\neg A\wedge \neg B$.
It is then possible to characterize all the possible values for $p$ by means of a linear system.
Notice that probabilistic logic by itself does not manipulate conditional probabilities or any notion of independence.
Proposals for extending the probabilistic logic to conditionals have appeared rather early~\cite{adams},
but Andersen and Hooker~\cite{BayesianLogic} introduced an efficient modeling and solve of such problems.
This new paradigm for manipulating Bayesian probabilistic constraints has been called \emph{Bayesian Logic}.
It is not linear.
For example, constraints like $p(A\wedge B)=p(A)p(B)$ remains essentially a non-linear constraint.
Constraints involving both conditional and non-conditional probabilities also generate non-linearity.
In~\cite{BayesianLogic}, Andersen and Hooker expose a methodology for solving these non-linear programs.
In particular, the structure of the Bayesian Network is being used in order to reduce the number of non-linear constraints.
\\[5pt]
\emph{Bayesian Logic} is a paradigm for solving probabilistic constraint programs, which involve Bayesian constraints.
Since it does not construct the Bayesian conditional as a strict logical operator, this theory is not concerned by Lewis' triviality.
\emph{Bayesian Logic} departs fundamentally from our approach, since \emph{Deterministic Bayesian Logic} intends to build the logic underlying the Bayesian conditional priorly to the notion of probability.
\paragraph{Conditional Event algebra.}
In Conditional Event Algebra~\cite{calabrese}, the conditional could be seen as an external operator $(\,|\,)$\,, which maps pairs of unconditioned propositions toward an \emph{external} Boolean space.
There are numerous possible constructions of a CEA.
Some typical properties related to the Bayesian conditional are generally implemented:
\begin{itemize}
\item Inference property: $(a|b\wedge c)\wedge (b|c)=(a\wedge b|c)\,,$
\item Boolean compatibility:
$(a\wedge b|c)=(a|c)\wedge(b|c)\quad\mbox{and}\quad(a\vee b|c)=(a|c)\vee(b|c)\,.$
\end{itemize}
But most CEAs provide conditional rules which are richer than the strict Bayesian conditional, and in particular compute the combination of any pairs of conditionals.
\\[5pt]
The counterpart of such nice properties is the necessity to restrict the conditional to unconditioned propositions.
The external space hypothesis is thus fundamental here.
CEAs are practically restricted to only one level of conditioning, and usually avoid any interferences between unconditioned and conditioned propositions.
These restrictions are the way, by which CEAs avoid Lewis' triviality.
\paragraph{Conditional logics.}
\emph{Conditional} is an ambiguous word, since there may be different meaning owing to the community.
Even the classical inference, $\phi\rightarrow\psi\equiv \neg\phi\vee\psi$\,, is called \emph{material conditional} by some.
Despite classical inference is systematically used by mathematicians, its disjunctive definition makes it improper for some conditions of use.
For example, it is known that it is by essence non-constructive, an issue which tracks back to the foundation of modern mathematics~\cite{intuitionism}.
\\[5pt]
Material conditional could be particularly improper for describing the logic of human mind.
For example, consider the sentences:
\begin{enumerate}
\item \label{dbl:sentence:1} ``If Robert were in Berlin, then he would be in France''\,,
\item \label{dbl:sentence:2} ``If Robert were in Berlin, then he would be in Germany''\,.
\end{enumerate}
Since Germany and France are two distinct countries, a human will say that sentence~\ref{dbl:sentence:1} is false, while sentence~\ref{dbl:sentence:2} is true.
For a human, moreover, the meaning of the sentences are \emph{independent} of the fact that Robert is in Berlin or not.
Now, interpreting~\ref{dbl:sentence:1} as a material conditional, it happens that this sentence is true when Robert is not in Berlin.
Sentences~\ref{dbl:sentence:1} and~\ref{dbl:sentence:2} should not be actually interpreted as material conditionals.
In fact, they are called \emph{Counterfactual conditionals}, and their truth does not depend on the truth of their hypotheses and conclusions.
The philosophers David Lewis and Robert Stalnaker have done fundamental works on counterfactual conditionals~\cite{Lewis:counterfactuals,stalnaker}.
While defining counterfactual conditionals
(an example of such conditional, VCU, is detailed in section~\ref{Section:Theorem:DBL:sub:2}), they based their model constructions on the \emph{possible world semantics} of modal logic.
Beside, couterfactuals and other related conditionals are deeply connected to the notion of logical modalities~\cite{giordano2}.
\\[5pt]
Actually, if we interpret the Bayesian conditional as a probabilistic conditional proposition, \emph{i.e.} $p(B|A)=p(A > B)$, it is derived $p(A)p(A > B)=p(A \wedge B)=p\bigl(A\wedge(A > B)\bigr)$, which means a probabilistic independence between $(A > B)$ and $A$.
So, it is tempting to interpret the Bayesian conditional as a counterfactual.
Stalnaker claimed that it was possible to construct such conditional within the universe of events, so as to match the Bayesian conditional.
Lewis answered negatively~\cite{lewis} to this conjecture.
Nevertheless, Lewis proposed an alternative interpretation of the probability $p(A > B)$ of the conterfactual $A>B$, called Imaging~\cite{lepage}, which give up the strong constraint $p(A > B)=p(B|A)$.
\\[5pt]
At last, it appears that (counterfactual) conditional logics are, in principle, a nice framework for interpreting the Bayesian inference, but the problem of the triviality have to be overcome.
As already explained in our previous discussion, the triviality should be avoided by constructing the conditionals outside the classical propositions space and extending the probability accordingly.
Now, existing conditional logics, often inspired from VCU, fail to implement some natural properties of the Bayesian conditional.
This is particularly true, when considering the negation of propositions.
Since the conditional probability of the negation is obtained as the complement of the conditional probability, \emph{i.e} $p(\sim B|A)=1-p(B|A)$, it seems natural to hypothesize a similar logical relation, \emph{e.g.} $(\neg\psi|\phi)\equiv\neg(\psi|\phi)$; \emph{notice that this relation is generally implemented by Conditional Event Algebras}.
This relation is not implemented by conditional logics in general (refer to the logic VCU defined in section~\ref{Section:Theorem:DBL:sub:2}).
In particular, the relation \mbox{$\neg(\phi>\psi)\equiv\phi>\neg\psi$} would contradict the axiom $\phi>\phi\mbox{ (Id)}$ which is widely accepted in the literature; refer to deduction (\ref{DBL:VCU:eq1}) in section~\ref{Section:Theorem:DBL:sub:2}.
\paragraph{Contribution.}
\emph{Bayesian logic} does not provide a logical interpretation of the Bayesian conditional, but rather a methodology for solving the program related to a probabilistic Bayesian modeling.
Existing \emph{conditional algebras} and \emph{conditional logics} are restricted or insufficient for characterizing the Bayesian conditional properly.
Our work intends to supply these limitations, by constructing a new conditional logic, denoted \emph{Deterministic Bayesian Logic} (DBL), which is in accordance with the Bayesian conditional.
The conditional operator is defined conjointly with a meta-relation of logical independence.
The probabilistic Bayesian inference is recovered from the derived logical theorems and the logical independence.
This process implies an extension of probability from the unconditioned logic toward DBL.
As a final result, a theorem is proved that guarantees the existence of such extension (Lewis result is thus avoided).
\\[5pt]
Section~\ref{Section:Def:DBL} is dedicated to the definition of the Deterministic  Bayesian Logic.
The languages, axioms and rules are introduced.
In section~\ref{Section:Theorem:DBL}, several theorems of the logic are derived.
A purely logical interpretation of Lewis' triviality is made, and DBL is compared with the known conditional logic VCU. 
A model for DBL is constructed in section~\ref{Section:Model:DBL}.
A completeness theorem is derived.
The extension of probabilities over DBL is investigated in section~\ref{Section:Proba:DBL}.
The probabilistic Bayesian inference is recovered from this extension.
The paper is then concluded.
\section{Definition of the logic}
\label{Section:Def:DBL}
The \emph{Deterministic Bayesian Logic} is defined now.
This definition implies a notion of logical independence, which is related to the proof of the propositions.
Typically, the following property holds true for the models of our logic:
\begin{equation}\label{eq:fond:0}\rien\qquad\begin{array}{@{}l@{}}
\mbox{Assume }\phi\mbox{ and }\psi\mbox{ to be logically independent.}\\
\mbox{Then }\phi\vee\psi\mbox{ is a tautology implies }\phi\mbox{ is a tautology or }\psi\mbox{ is a tautology}\,.
\end{array}\end{equation}
In this document, we propose a definition based on the \emph{sequent} formalism (there is a previous modal embedded definition~\cite{Dambreville:DmBL}). 
However, although the definition is formalized by means of sequent, it does not retain the rules of sequent calculus~\cite{girard}.
For a soft introduction of the logic, informal intuitions about the Bayesian inference are given now.
\subsection{Logical relations within Bayesian probability}
\label{Section:Def:DBL:intro}
Here, some typical probabilistic relations are considered, and logical theorems
and axioms are extrapolated from these relations.
{\bf These extrapolations are not justified here;} it is the purpose of the paper to prove the coherence of the whole logic, while this paragraph is only dedicated to the intuitions behind the formalism.
\\[5pt]
The logic of a system may be seen as the collection of behaviors which are common to any instance of this system.
Let us consider the example of probability on a finite \emph{(unconditioned)} propositional space.
For convenience, define $\mathbb{P}$ the set of strictly positive probabilities over this space, that is $p\in\mathbb{P}$ is such that $p(\phi)>0$ for any non-empty proposition~$\phi$\,:
$$
\mathbb{P}=\bigl\{p\;\big/\;p\mbox{ is a probability and }\forall\phi\not\equiv\bot,\,p(\phi)>0\bigr\}\,.
$$
Then, the following properties are easily derived for \emph{unconditioned} propositions:
\begin{eqnarray}
&&\label{JYG:1:1} \forall p\in\mathbb{P},\,p(\phi)+p(\psi)=1\quad\mbox{implies}\quad\phi\equiv\neg\psi\;,
\\
&&\label{JYG:1:2} \forall p\in\mathbb{P},\,p(\phi)+p(\psi)\le p(\eta)+p(\zeta)\quad\mbox{implies}\quad\vdash(\phi\vee\psi)\rightarrow(\eta\vee\zeta)\;,
\end{eqnarray}
with corollary:
\begin{equation}
\label{JYG:1:3}\forall p\in\mathbb{P},\,p(\phi)=1\quad\mbox{implies}\quad\vdash\phi\;.
\end{equation}
For any $p\in\mathbb{P}$, define the conditional extension $\overline{p}$ by:
$$
\overline{p}(\psi|\phi)p(\phi)=p(\phi\wedge\psi)\;,
\quad\mbox{for any unconditioned propositions } \phi,\psi\,.
$$
While results~(\ref{JYG:1:1}) to~(\ref{JYG:1:3}) work for unconditioned propositions, we extrapolate them to some elementary conditional relations related to $\overline{p}$:
\begin{itemize}
\item It is noticed that $\forall p\in\mathbb{P},\;\overline{p}(\psi|\phi)+\overline{p}(\neg\psi|\phi)=1$\,.
Property~(\ref{JYG:1:1}) could be extrapolated to $\overline{p}(\psi|\phi)$ and $\overline{p}(\neg\psi|\phi)$, and then yields:
$$
\neg(\psi|\phi)\equiv(\neg\psi|\phi)\;.
$$
Of course, although intuitively sound, this relation is not justified mathematically.
This logical relation is implemented in DBL as the axiom b4, and is expressed as a sequent:
\begin{equation}\label{JYG:2:1}
\vdash\neg(\psi|\phi)\leftrightarrow(\neg\psi|\phi)\;.
\end{equation}
\item 
It is known that $\forall p\in\mathbb{P},\;\overline{p}(\psi|\phi)+p(\bot)\le p(\neg\phi\vee\psi)+p(\bot)$\,.
Then, the extrapolation of property~(\ref{JYG:1:2}) yields:
\begin{equation}
\label{JYG:2:2}
\vdash(\psi|\phi)\rightarrow(\phi\rightarrow\psi)\;.
\end{equation}
This logical relation is implemented in DBL as the axiom b3.
\item
Similarly, it comes $\forall p\in\mathbb{P},\;\overline{p}(\psi\vee\eta|\phi)+p(\bot)\le \overline{p}(\psi|\phi)+\overline{p}(\eta|\phi)$.
Then, the extrapolation of~(\ref{JYG:1:2}) yields:
$$
\vdash(\psi\vee\eta|\phi)\rightarrow\bigl((\psi|\phi)\vee(\eta|\phi)\bigr)\,.
$$
Together with~(\ref{JYG:2:1}), it is then deduced:
\begin{equation}
\label{JYG:2:3}
\vdash(\psi\rightarrow\eta|\phi)\rightarrow\bigl((\psi|\phi)\rightarrow(\eta|\phi)\bigr)\;,
\end{equation}
which constitutes a modus ponens for the conditional.
This logical relation is implemented in DBL as the axiom b2.
\item Another interesting relation is: 
$$
\vdash\phi\rightarrow\psi\mbox{ implies }\vdash\neg\phi\mbox{ or }\forall p\in\mathbb{P},\;\overline{p}(\psi|\phi)=1\;.
$$
Extrapolating~(\ref{JYG:1:3}), it comes:
$$
\vdash\phi\rightarrow\psi\mbox{ implies }\vdash\neg\phi\mbox{ or }\vdash(\psi|\phi)\;.
$$ 
This logical relation is implemented in DBL as the axiom b4, and is expressed as a sequent:
\begin{equation}\label{JYG:2:4}
\phi\rightarrow\psi\vdash\neg\phi,(\psi|\phi)\;.
\end{equation}
\emph{Notice that in DBL, sequent $\vdash\phi,\psi$ is not equivalent to $\vdash\phi\vee\psi$\,.}
In fact, this axiom is related to the property~(\ref{eq:fond:0})\,, mentioned previously.
\item From the Bayesian inference, it is known that $\overline{p}(\psi|\phi)=\overline{p}(\psi)$ implies $\overline{p}(\phi|\psi)=\overline{p}(\phi)$\,.
By similar extrapolation, it is then derived:
$$\vdash(\psi|\phi)\leftrightarrow\psi\mbox{ implies }\vdash(\phi|\psi)\leftrightarrow\phi\,,$$
Notice however that this relation does not make sense in general, when $\phi$ and $\psi$ are both unconditioned propositions.
This logical relation is implemented in DBL as the axiom b5, and is expressed as a sequent:
\begin{equation}\label{JYG:2:5}
\psi\times\phi\vdash\phi\times\psi\,,
\mbox{ where }
\phi\times\psi=(\phi|\psi)\leftrightarrow\phi\,,
\mbox{ and }
\psi\times\phi=(\psi|\phi)\leftrightarrow\psi\,.
\end{equation}
\end{itemize}
In fact, these extrapolated axioms imply constraints, when extending the probabilities $p\in\mathbb{P}$ over the conditioned propositions.
This paper intends to prove that these constraints are actually valid, in regard to Lewis' triviality.
It is now time for the logic definition.
\subsection{Language}
Let $\Theta=\{\theta_i/i\in I\}$ be a \emph{finite} set of atomic propositions.
\\[5pt]
The language $\mathcal{L}_C$ of the classical logic related to $\Theta$ is the smallest set such that:
$$\left\{\begin{array}{@{\,}l@{}}
\Theta\subset\mathcal{L}_C
\\[5pt]
\neg\phi\in\mathcal{L}_C \mbox{ and }\phi\rightarrow\psi\in\mathcal{L}_C
\mbox{ for any }\phi,\psi\in\mathcal{L}_C
\end{array}\right.$$
The language $\mathcal{L}$ of the D\emph{eterministic} B\emph{ayesian} L\emph{ogic} related to $\Theta$ is the smallest set such that:
$$\left\{\begin{array}{@{\,}l@{}}
\Theta\subset\mathcal{L}
\\[5pt]
\neg\phi\in\mathcal{L}\;,\ \phi\rightarrow\psi\in\mathcal{L}\mbox{ and }(\psi|\phi)\in\mathcal{L}
\mbox{ for any }\phi,\psi\in\mathcal{L}
\end{array}\right.$$
The following abbreviations are defined:
\begin{itemize}
\item
$\phi\vee\psi=\neg\phi\rightarrow\psi$\,,\quad $\phi\wedge\psi=\neg(\neg\phi\vee\neg\psi)$\quad
and\quad
$\phi\leftrightarrow\psi=(\phi\rightarrow\psi)\wedge(\psi\rightarrow\phi)$\,,
\item $\psi\times\phi=(\psi|\phi)\leftrightarrow\psi$\,,
\item It is chosen a proposition $\theta_1\in\Theta$, and it is then denoted $\top=\theta_1\rightarrow\theta_1$ and $\bot=\neg\top$\,.
\end{itemize}
The operator $\times$ is involved (subsequently) in the definition of the logical independence, though it is not sufficient to characterize this meta-relation by itself.
$\top$ and $\bot$ are idealistic notations for the tautology and the contradiction.
\subsection{Sequents}
The set of finite sequences of propositions of $\mathcal{L}$, denoted $\mathcal{L}^\ast$, is defined by:
\begin{equation}\label{DBL:def:sequence:eq1}
\mathcal{L}^\ast=\bigcup_{n=0}^{\infty}\mathcal{L}^n\,,
\end{equation}
where $\mathcal{L}^n$ is the set of n-uplets of $\mathcal{L}$.
In particular, $\mathcal{L}^0=\{\emptyset\}$ where $\emptyset$ is the empty sequence.
\paragraph{Notations.}
Subsequently, finite sequences of $\mathcal{L}^\ast$ are denoted without brackets.
\\[5pt]
Being given a finite sequence $\Gamma=\gamma_1,\dots,\gamma_n$ of $\mathcal{L}^\ast$, then $\{\Gamma\}=\{\gamma_1,\dots,\gamma_n\}$ is the set of all components of the sequence $\Gamma$\,.
Notice that the set $\{\Gamma\}$ may contain less components than the sequence $\Gamma$, since a sequence may repeat the same proposition.
\\[5pt]
Let $\Gamma=\gamma_1,\dots,\gamma_n$ and $\Delta=\delta_1,\dots,\delta_m$ be two finite sequences of $\mathcal{L}^\ast$\,.
Then $\Gamma,\Delta$ is the sequence $\gamma_1,\dots,\gamma_n\;,\;\delta_1,\dots,\delta_m$\,, obtained as a concatenation of $\Gamma$ and $\Delta$.
\paragraph{Definition.}
The set of sequents of $\mathcal{L}$, denoted $\seqset$, is defined as the set of pairs of finite sequences of $\mathcal{L}$: 
\begin{equation}\label{DBL:def:sequence:eq2}
\seqset=\mathcal{L}^\ast\times\mathcal{L}^\ast\,.
\end{equation}
\paragraph{Notation.}
Being given a subset $X\subset\seqset$ of sequents and a sequent $(\Gamma,\Delta)\in \seqset$, the meta-relation $\Gamma\vdash_X\Delta$ is defined by:
\begin{equation}\label{DBL:def:sequence:eq3}
\Gamma\vdash_X\Delta
\mbox{ if and only if }
(\Gamma,\Delta)\in X\;.
\end{equation}
When $\Gamma=\emptyset$ (resp. $\Delta=\emptyset$), the notation $\vdash_X\Delta$ (resp. $\Gamma\vdash_X$) is used instead of $\Gamma\vdash_X\Delta$.
\\[5pt]
Subsequently are defined the set of sequents deducible in DBL, denoted $\mathcal{B}$, and the set of sequents deducible classically, denoted $\mathcal{C}$.
These sets are defined by means of rules and axioms of construction.
Such axiomatic constructions depart from common sequent calculi, like LK.
\subsection{Rules and axioms}
The sets $\mathcal{B}\subset\seqset$, $\mathcal{B}_\ast\subset\seqset$ and  $\mathcal{C}\subset\seqset$ are defined as the smallest subsets of $\seqset$ verifying:
\begin{itemize}
\item
For $X\in\{\mathcal{B},\mathcal{B}_\ast,\mathcal{C}\}$:
\begin{description}
\item[CUT.] $\Gamma\vdash_X\Delta,\phi$ and $\Lambda,\phi\vdash_X\Sigma$ implies $\Gamma,\Lambda\vdash_X\Delta,\Sigma$\,,
\item[STRUCT.] 
Assume $\{\Gamma\}\subset\{\Lambda\}\cup\{\top\}$ and $\{\Delta\}\subset\{\Sigma\}\cup\{\bot\}$\,.\\
Then $\Gamma\vdash_X\Delta$ implies $\Lambda\vdash_X\Sigma$\,.
\item[Modus ponens.]$\phi,\phi\rightarrow\psi\vdash_X\psi$\,,
\item[Classical Axioms:]
\item[c1.] $\vdash_X \phi\rightarrow(\psi\rightarrow\phi)$\,,
\item[c2.] $\vdash_X (\eta\rightarrow(\phi\rightarrow\psi))\rightarrow((\eta\rightarrow\phi)\rightarrow(\eta\rightarrow\psi))$\,,
\item[c3.] $\vdash_X (\neg\phi\rightarrow\neg\psi)\rightarrow((\neg\phi\rightarrow\psi)\rightarrow\phi)$\,,
\end{description}
\item For $X\in\{\mathcal{B},\mathcal{B}_\ast\}$:
\begin{description}
\item[b1.] $\phi\rightarrow\psi\vdash_X\neg\phi,(\psi|\phi)$\,,
\item[b2.] $\vdash_X(\psi\rightarrow\eta|\phi)\rightarrow\bigl((\psi|\phi)\rightarrow(\eta|\phi)\bigr)$\,,
\item[b3.] $\vdash_X(\psi|\phi)\rightarrow(\phi\rightarrow\psi)$\,,
\item[b4.] $\vdash_X\neg(\neg\psi|\phi)\leftrightarrow(\psi|\phi)$\,,
\end{description}
\item For $X=\mathcal{B}$:
\begin{description}
\item[b5.] \emph{(logical independence is symmetric)}~:
$\psi\times\phi\vdash_X\phi\times\psi$\,,
\end{description}
\item For $X=\mathcal{B}_\ast$:
\begin{description}
\item[b5.weak.A.]
$\psi\times\neg\phi\vdash_X\psi\times\phi$
 and 
$\psi\times\phi\vdash_X\psi\times\neg\phi$\,,
\item[b5.weak.B.] $\psi\leftrightarrow\eta\vdash_X(\phi|\psi)\leftrightarrow(\phi|\eta)$\,.
\end{description}
\end{itemize}
$\mathcal{B}$ is the set of sequents deducible in DBL.
$\mathcal{C}$ is the set of sequents deducible classically.
The axioms b5.weak.A and b5.weak.B are actually a weakening of b5 (refer to section~\ref{Section:Theorem:DBL}).
The set $\mathcal{B}_\ast$ is thus related to a weakened version of DBL, denoted DBL$_\ast$.
In fact, DBL$_\ast$ is a quite useful intermediate  for the construction of a model of DBL.
It happens that the model of DBL$_\ast$ is constructed directly, while the model of DBL is implied from the model of DBL$_\ast$.
\paragraph{Notations.}
The following meta-abbreviations are defined for $X\in\{\mathcal{C},\mathcal{B},\mathcal{B}_\ast\}$\,:
\begin{itemize}
\item $\phi\equiv_X\psi$ means $\vdash_X\phi\leftrightarrow\psi$.
\end{itemize}
The relation $\equiv_X$ is the logical equivalence related to the deduction system $X$.
\\[5pt]
In order to alleviate the notations, the subscripts $_{\mathcal{B}}$ and $_{\mathcal{B}_\ast}$ are omitted.
In particular, $\vdash$ (resp. $\equiv$) is used instead of $\vdash_{\mathcal{B}}$ or $\vdash_{\mathcal{B}_\ast}$ (resp. $\equiv_{\mathcal{B}}$ or $\equiv_{\mathcal{B}_\ast}$).
\paragraph{Notations relative to $(\cdot|\cdot)$.}
The set $\bigl\{
\eta\in\mathcal{L}
\;\big/\;
\exists\psi\in\mathcal{L},\,\eta\equiv(\psi|\phi)
\bigr\}$ is called the \emph{sub-universe} of $\phi$.
\\[5pt]
The \emph{logical} independence between propositions is a meta-relation expressed from the operator $\times$ and the sequents:
\begin{center}
By definifion, $\psi$ is logically independent of $\phi$\,, when $\vdash\psi\times\phi$.
\end{center}
The logical independence and the conditional $(|)$ are thus conjointly defined.
\paragraph{Interpretation.}
The construction of the model in section~\ref{Section:Model:DBL} infers the following interpretation of sequent $\Gamma\vdash\Delta$:
\begin{equation}\label{DBL:def:Sequent:interpret:eq1}\rien\hspace{40pt}\begin{array}{@{}l@{}}
\mbox{If all propositions }\gamma\in\{\Gamma\}\mbox{ are tautologies of the model,}
\\\mbox{then there is a proposition }\delta\in\{\Delta\}\mbox{ which is a tautology of the model.}
\end{array}\end{equation}
\paragraph{Meaning of the rules and axioms.}
Axioms $c\ast$ are well known minimal axioms of classical logic.
Axioms $b\ast$ have been introduced in section~\ref{Section:Def:DBL:intro}, and are thought to describe the logical behavior of a Bayesian operator.
Axiom \emph{modus ponens} is the modus ponens rule encoded within a sequent formalism.
Rule CUT is the well known \emph{cut} rule for merging sequent proofs.
Rule STRUCT is a structural rule for the sequent, which includes the weakening, contraction and permutation.
Moreover, it makes possible to suppress $\top$ (resp. $\bot$) from the left (resp. right) side of a sequent.
In particular, STRUCT makes equivalent $\Gamma\vdash\Delta$ and  $\Gamma,\top\vdash\Delta,\bot$.
\subsection{DBL extends classical logic.}
It is noticed that the \emph{classical Logic $C$} is obtained by restricting DBL to the language $\mathcal{L}_C$ and to the deduction rules CUT and STRUCT, the axiom \emph{modus ponens} and the classical axioms $c\ast$ described previously (\emph{c.f.} appendix~\ref{proof:clasrestric}).
More precisely, if $\phi$ is a theorem of classical logic, then it is deduced $\vdash_C\phi$.
So, in some common sense, DBL extends classical logic.
However, \emph{the rules of LK} (a common sequent calculus for classical logic) \emph{do not work anymore in our system}, and moreover, there are sequents deduced from LK which cannot be derived from our deduction system.
Examples are provided in appendix~\ref{proof:clasrestric}.
Thus, one have to be careful in the deduction process of classical sequents.
\\[5pt]
While DBL could be seen as an extension of $C$, the following properties are desirable:
\begin{itemize}
\item If $\phi\in\mathcal{L}_C$, then $\vdash \phi$ implies $\vdash_C \phi$\,,
\item For any probability $p$ defined over $\mathcal{L}_C$\,, there is a probability $\overline{p}$ over $\mathcal{L}$ which extends $p$ and verifies $\overline{p}(\phi\wedge\psi)=\overline{p}(\phi)\overline{p}(\psi)$, for any $\phi,\psi\in\mathcal{L}$ such that $\vdash\psi\times\phi$ (logical independence)\,.
\end{itemize}
First property just ensures that DBL axioms will not trivialize the classical logic.
Second property ensures that DBL is not just a trivial extension of $C$, and in particular avoids the triviality of Lewis.
These results are amongst the most contributions of this paper.
Another main contribution of the paper is that such extention $\overline{p}$ actually implies the probabilistic Bayesian inference:
$$
\overline{p}\bigl((\psi|\phi)\bigr)p(\phi)=p(\phi\wedge\psi)\,,\mbox{ for any }\phi,\psi\in\mathcal{L}_C\,.
$$
These results are obtained from the model constructed for DBL.
But first, the following section studies the logical consequences of the rules and axioms of DBL.
\section{Logical theorems and comparison}
\label{Section:Theorem:DBL}
Subsequently, theorems of DBL are derived.
Since both DBL and DBL$_\ast$ are studied, \emph{the possibly needed axioms $b5\ast$ are indicated in bracket.}
\\[5pt]
First at all, it happens that both DBL and DBL$_\ast$ imply the classical tautologies.
In particular, the following property is proved in appendix~\ref{proof:clasrestric}:
\begin{center}
Assume that $\phi$ is a tautology of classical logic.
Then $\vdash_C\phi$ is deduced from the classical subsystem of DBL.
\end{center}
For this reason, the theorems of classical logic are assumed without proof from now on, so that many details in the deductions are implied.
\subsection{Theorems}
\label{Section:Theorem:DBL:sub:1}
\emph{The proofs are done in appendix~\ref{proof:logth}.}
Next theorem is proved here as an example. 
%
\subsubsection{The full universe}\label{DBL:theo:1} $\phi\vdash\psi\times\phi$\,.
In particular $(\psi|\top)\equiv\psi$\,.\\[5pt]
Interpretation: a tautology is independent with any other proposition and its sub-universe is the whole universe.
\begin{description}
\item[Proof.]
From axiom b3, it comes $\vdash(\psi|\phi)\rightarrow(\phi\rightarrow\psi)$ and  $\vdash(\neg\psi|\phi)\rightarrow(\phi\rightarrow\neg\psi)$\,.\\
Then $\vdash\phi\rightarrow\bigl((\psi|\phi)\rightarrow\psi\bigr)$ and  $\vdash\phi\rightarrow\bigl((\neg\psi|\phi)\rightarrow\neg\psi\bigr)$\,, by classical deductions.\\
Applying b4 (with CUT and classical deductions) yields $\vdash\phi\rightarrow\bigl((\psi|\phi)\leftrightarrow\psi\bigr)$.\\
By applying modus ponens and CUT, it follows $\phi\vdash(\psi|\phi)\leftrightarrow\psi$\,.
\\[5pt]
The remaining proof is obvious.
\item[$\Box\Box\Box$]\rien
\end{description}
\subsubsection{Axioms order}\label{DBL:theo:2} Axiom b5 implies b5.weak.A.
\subsubsection{The empty universe [needs b5.weak.A]}\label{DBL:theo:3} $\neg\phi\vdash\psi\times\phi$\,.
In particular $(\psi|\bot)\equiv\psi$\,.
\subsubsection{Left equivalences}\label{DBL:theo:4}
$\psi\leftrightarrow\eta\vdash\neg\phi,(\psi|\phi)\leftrightarrow(\eta|\phi)$\,.
\\[5pt]
\emph{Corollary [b5.weak.A].} $\psi\leftrightarrow\eta\vdash(\psi|\phi)\leftrightarrow(\eta|\phi)$.
\\[5pt]
\emph{Corollary 2 [b5.weak.A].} $\psi\equiv\eta$ implies $(\psi|\phi)\equiv(\eta|\phi)$.
\\
Proof of corollary~2 is immediate from corollary.
\subsubsection{Sub-universes are classical [b5.weak.A]}\label{DBL:theo:5}
\begin{itemize}
\item $(\neg\psi|\phi)\equiv\neg(\psi|\phi)$\,,
\item $(\psi\wedge\eta|\phi)\equiv(\psi|\phi)\wedge(\eta|\phi)$\,,
\item $(\psi\vee\eta|\phi)\equiv(\psi|\phi)\vee(\eta|\phi)$\,,
\item $(\psi\rightarrow\eta|\phi)\equiv(\psi|\phi)\rightarrow(\eta|\phi)$\,.
\end{itemize}
\subsubsection{Evaluating $(\top|\cdot)$ and $(\bot|\cdot)$ [b5.weak.A]}\label{DBL:theo:6} $\psi\vdash(\psi|\phi)$\,.
In particular $(\top|\phi)\equiv\top$ and $(\bot|\phi)\equiv\bot$\,.
\subsubsection{Inference property}\label{DBL:theo:7}
$(\psi|\phi)\wedge\phi\equiv\phi\wedge\psi$\,.
\\[5pt]
Interpretation: the Bayesian conditional is actually an inference.
\subsubsection{Introspection}\label{DBL:theo:8} $\vdash\neg\phi,(\phi|\phi)$\,.\\[5pt]
Interpretation: a non-empty proposition sees itself as ever true. \\
Notice that this property is compliant with $(\bot|\bot)\equiv\bot$\,, itself derived from~\ref{DBL:theo:3}.
\subsubsection{Inter-independence [b5.weak.A]}\label{DBL:theo:9} $\vdash(\psi|\phi)\times\phi$\,.\\[5pt]
Interpretation: a proposition is independent of its sub-universe.
\subsubsection{Independence invariance [b5.weak.A]}\label{DBL:theo:10}
$$
\begin{array}{@{}l@{}}
\psi\times\phi \vdash \neg\psi\times\phi\;,
\\
\psi\times\phi,\eta\times\phi\vdash(\psi\wedge\eta)\times\phi\;,
\\
\psi\leftrightarrow\eta,\psi\times\phi\vdash\eta\times\phi\;.
\end{array}
$$
\subsubsection{Narcissistic independence}\label{DBL:theo:11} $\phi\times\phi\vdash\neg\phi,\phi$\,.
\\[5pt]
Interpretation: a propositions independent with itself is either a tautology or a contradiction.
\subsubsection{Independence and proof [b5.weak.A]}\label{DBL:theo:12}
$\psi\times\phi,\phi\vee\psi\vdash\phi,\psi$\,.\\[5pt]
Interpretation:
when propositions are independent and their disjunctions are proved, then at least one proposition is ``proved''.
\subsubsection{Independence and regularity [b5.weak.A]}\label{DBL:theo:13}
$$
\phi\times\eta,\psi\times\eta,(\phi\wedge\eta)\rightarrow(\psi\wedge\eta)\vdash\neg\eta,\phi\rightarrow\psi\;.
$$
Interpretation: unless it is empty, a proposition may be removed from a logical equation, when it appears in the both sides and is independent with the equation components.
\\[5pt]
\emph{Corollary.} $\vdash\phi\times\eta$\,, $\vdash\psi\times\eta$\,, $\neg\eta\vdash$ and $\phi\wedge\eta\equiv\psi\wedge\eta$ imply $\phi\equiv\psi$\,.
\\[5pt]
\emph{Corollary 2.} Being given $\psi$ and $\phi$ such that $\neg\phi\vdash$, then $(\psi|\phi)$ is uniquely defined as the solution of equation $X\wedge\phi\equiv\psi\wedge\phi$ (with unknown $X$) which is independent of $\phi$.
%
\subsubsection{Right equivalences [b5]}\label{DBL:theo:14}
$\psi\leftrightarrow\eta\vdash(\phi|\psi)\leftrightarrow(\phi|\eta)$ (proved with b5 but without b5.weak.B).
\\[5pt]
Interpretation: equivalence is compliant with the conditioning.
\\[5pt]
\emph{Corollary.} Axiom b5 implies b5.weak.B.
In particular, DBL$_\ast$ is weaker than DBL.
\\[5pt] 
\emph{Corollary of b5 or b5.weak.B.} $\psi\equiv\eta$ implies $(\phi|\psi)\equiv(\phi|\eta)$.
\\[5pt]
Combined with the classical theorems and~\ref{DBL:theo:4}, this last result implies that the equivalence relation $\equiv$ is compliant with the logical operators of DBL/DBL$_\ast$.
In particular, replacing a sub-proposition with an equivalent sub-proposition within a theorem still makes a theorem.
\subsubsection{Reduction rule [b5]}\label{DBL:theo:15}
Axiom b5 implies $\bigl(\phi\big|(\psi|\phi)\bigr)\equiv\phi$\,.
\subsubsection{Markov Property [b5]}\label{DBL:theo:16}
$$
(\phi_t|\phi_{t-1})\times\phi_1,\dots,(\phi_t|\phi_{t-1})\times\phi_{t-2}
\ \vdash\ 
\neg\left(\bigwedge_{\tau=1}^{t-1}\phi_{\tau}\right)\;,\;
(\phi_t|\phi_{t-1})\leftrightarrow\left(\phi_t\left|\bigwedge_{\tau=1}^{t-1}\phi_{\tau}\right.\right)
\;.
$$
Interpretation: the Markov property holds, when the conditioning is independent of the past and the past is possible.
\subsubsection{Link between $\bigl((\eta|\psi)\big|\phi\bigr)$ and $(\eta|\phi\wedge\psi)$ [b5]}\label{DBL:theo:17}
It is derived:
$
\bigl((\eta|\psi)\big|\phi\bigr)\wedge\phi\wedge\psi\equiv(\eta|\psi)\wedge\phi\wedge\psi\equiv\phi\wedge\psi\wedge\eta\equiv(\eta|\phi\wedge\psi)\wedge(\phi\wedge\psi)\;.
$\\[5pt]
This is a quite limited result and it is \emph{tempting} to hypothesize the additional axiom ``$\bigl((\eta|\psi)\big|\phi\bigr)\equiv(\eta|\phi\wedge\psi)\quad\mbox{\small$(\ast)$}$''\,.
There is a really critical point here, since axiom~$(\ast)$ implies actually a logical counterpart to Lewis' triviality\,:
\begin{quote}
\emph{Let $\bigl((\eta|\psi)\big|\phi\bigr)\equiv(\eta|\phi\wedge\psi)\quad\mbox{\small$(\ast)$}$ be assumed as an axiom.\\
Then $\vdash\neg(\phi\wedge\psi),\phi\leftrightarrow\psi,\phi\times\psi$\,.}
\end{quote}
Interpretation: if $\phi$ and $\psi$ are not exclusive and not equivalent, then they are independent.
This is irrelevant and forbids the use of axiom~$(\ast)$.
\subsection{Comparison with the conditional logic VCU}
\label{Section:Theorem:DBL:sub:2}
The axioms of the conditional logic VCU (VCU is an abbreviation for the axioms system)~\cite{Lewis:counterfactuals} are considered here and compared to DBL.
This example is representative of the difference with some other conditional logics.
\emph{Theorems derived in section~\ref{Section:Theorem:DBL:sub:1} are referred to.}
\\[5pt]
The language of VCU involves a counterfactual inference operator $\counterfact$ in addition to the classical operators.
This operator is characterized by the axioms Ax.1~to~Ax.6 and the counterfactual rule CR expressed as follows:
\begin{description}
\item[(Ax.1)] $\phi\counterfact\phi$\,,
\item[(Ax.2)] $(\neg\phi\counterfact\phi)\rightarrow(\psi\counterfact\phi)$\,,
\item[(Ax.3)] $(\phi\counterfact\neg\psi)\vee(((\phi\wedge\psi)\counterfact\xi)\leftrightarrow(\phi\counterfact(\psi\rightarrow\xi)))$\,,
\item[(Ax.4)] $(\phi\counterfact\psi)\rightarrow(\phi\rightarrow\psi)$\,,
\item[(Ax.5)] $(\phi\wedge\psi)\rightarrow(\phi\counterfact\psi)$\,,
\item[(Ax.6)] $(\neg\phi\counterfact\phi)\rightarrow\bigl(\neg(\neg\phi\counterfact\phi)\counterfact(\neg\phi\counterfact\phi)\bigr)$\,,
\item[(CR)] Being proved $(\xi_1\wedge\dots\wedge\xi_n)\rightarrow\psi$\,, it is proved $((\phi\counterfact\xi_1)\wedge\dots\wedge(\phi\counterfact\xi_n))\rightarrow(\phi\counterfact\psi)$\,.
\end{description}
It appears that Ax.2, Ax.4, Ax.5, Ax.6 and CR are recovered in DBL.
More precisely, Ax.2 becomes $\vdash(\phi|\neg\phi)\rightarrow(\phi|\psi)$ (derived from theorems).
Ax.4 is exactly b3.
Ax.5 is a subcase of $\phi\wedge\psi\equiv\phi\wedge(\psi|\phi)$ (inference theorem).
Ax.6 becomes
$\vdash(\phi|\neg\phi)\rightarrow\bigl((\phi|\neg\phi)\big|\neg(\phi|\neg\phi)\bigr)$
(derived from theorems).
And CR is recovered in DBL from the fact that \emph{sub-universes are classical}:
$$
\vdash(\xi_1\wedge\dots\wedge\xi_n)\rightarrow\psi\mbox{ implies }
\vdash((\xi_1|\phi)\wedge\dots\wedge(\xi_n|\phi))\rightarrow(\psi|\phi)\,.
$$
Ax.1 has a partial counterpart in DBL, \emph{i.e.} $\vdash\neg\phi,(\phi|\phi)$ (theorem).
However Ax.3 has no obvious counterpart in DBL.
\\[10pt]
Conversely, b3 is clearly implemented by VCU.
It is also noteworthy that Ax.1 and CR, with $n=1$ and $\xi_1=\phi$, infer the rule:
\begin{equation}
\label{DBL:VCU:1}
\mbox{Being proved }\phi\rightarrow\psi,\mbox{ it is proved }\phi\counterfact\psi\;,
\end{equation}
which is stronger than b1.
Although b2 is not implemented by VCU, it is easily shown that VCU completed by b4 implies b2.
The fact is that b4 is not implemented by VCU.
Moreover, b5 is related to the notion of logical independence, which is not considered within VCU.
\\[5pt]
Then we have to point out three fundamental distinctions of DBL compared to VCU:
\begin{enumerate}
\item\label{DBL:Point:1} In DBL, the negation commutes with the conditional (b4).
More generally, sub-universes are classical in DBL,
\item\label{DBL:Point:2} In DBL, the deductions on the conditionals are often weakened by the hypothesis that \emph{the condition is not empty}; for example, $\neg\phi$ in rule b1, or theorem $\vdash(\phi|\phi),\neg\phi$\,,
\item DBL manipulates a notion of logical independence of the propositions.
\end{enumerate}
\emph{In fact, point~\ref{DBL:Point:1} (commutation of the negation) makes point~\ref{DBL:Point:2} (deduction weakened by the non-empty condition hypothesis) necessary.}
For example, $\bot\counterfact\top$ is derived from (\ref{DBL:VCU:1});
by using both Ax.1 and the negation commutation, it is then deduced:
\begin{equation} \label{DBL:VCU:eq1}
\top\equiv\bot\counterfact\bot\equiv\bot\counterfact\neg\top\equiv\neg(\bot\counterfact\top)\equiv\neg\top\equiv\bot\;,
\end{equation}
which is impossible.
Notice that this deduction is also done in DBL, if we replace the ``weakened'' theorem $\vdash\neg\phi,(\phi|\phi)$ by the ``strong'' theorem $\vdash(\phi|\phi)$\,.
\\[5pt]
This example, based on VCU and DBL, illustrates a fundamental difference between DBL and other conditional logics.
DBL considers $\bot$ as a singularity, and will be cautious with this case when inferring conditionals.
This principle is not just a logical artifact.
In fact, it is also deeply related to the notion of logical independence, as it appears in the proof of theorem~\ref{DBL:theo:12}, \emph{Independence and proof}.
\section{Models}
\label{Section:Model:DBL}
\subsection{Definitions}
\paragraph{Notations of Boolean algebra.}
Being given a Boolean algebra~\cite{Boolean:algebra}, $(\mathbf{B},\cup,\cap,\sim,\emptyset,\Omega)$, the binary operators $\cup$ and $\cap$ are respectively the Boolean addition and multiplication, the unary operator $\sim$ is the Boolean complementation, $\emptyset$ and $\Omega$ are the neutal element for $\cup$ and $\cap$ respectively.
Moreover, the order $\subset$ is defined over $\mathbf{B}$ by setting for any $A,B\in \mathbf{B}$:
\begin{center}
$A\subset B$ if and only if $A\cap B=A\;.$
\end{center}
\paragraph{Definition of a conditional model.} A conditional model for DBL (respectively DBL$_\ast$) is a septuplet $\mathbf{M}=(\mathbf{B},\cup,\cap,\sim,\emptyset,\Omega,f)$, where $(\mathbf{B},\cup,\cap,\sim,\emptyset,\Omega)$ is a Boolean algebra, $f:\mathbf{B}\times \mathbf{B}\longrightarrow \mathbf{B}$, and verifying for any $A,B,C\in \mathbf{B}$:
\begin{description}
\item[$\rien\quad\beta1$.] $A\subset B$ and $A\ne\emptyset$ imply $f(B,A)=\Omega$\,,
\item[$\rien\quad\beta2$.] $f(B\cup C,A)\subset f(B,A)\cup f(C,A)$\,,
\item[$\rien\quad\beta3$.] $A\cap f(B,A) \subset B$\,,
\item[$\rien\quad\beta4$.] $f(\sim B,A)=\sim f(B,A)$\,,
\item[$\rien\quad\beta5$.] $f(B,A)=B$ implies $f(A,B)=A$\,,\\
(respectively $\beta5w.$ $f(B,A)=B$ implies  $f(B,\sim A)=B$)\,.
\end{description}
The objects $\cup,\cap,\sim,\emptyset,\Omega,f$ are a model conterpart of the logical objects $\vee,\wedge,\neg,\bot,\top,(\cdot|\cdot)$.
\paragraph{Definition of a conditional assignment.}
Let $\mathbf{M}=(\mathbf{B},\cup,\cap,\sim,\emptyset,\Omega,f)$ be a conditional model.
\\[5pt]
An atomic assignment on $\mathbf{M}$ is a mapping $h:\Theta\rightarrow \mathbf{B}$\,.
\\[5pt]
A conditional assignment on  $\mathbf{M}$ is a mapping $H:\mathcal{L}\rightarrow \mathbf{B}$ such that:
\begin{itemize}
\item $H(\neg\phi)=\sim H(\phi)$,
\item $H(\phi\rightarrow\psi)=\sim H(\phi)\cup H(\psi)$,
\item $H\bigl((\psi|\phi)\bigr)=f(H(\psi),H(\phi))$
\end{itemize}
for any $\phi,\psi\in\mathcal{L}$\,.
\\[5pt]
The set of all conditional assignment on $\mathbf{M}$ is denoted $\mathcal{H}[\mathbf{M}]$\,.
\begin{proposition}\label{DBL:prop:condass:1}
Let $h$ be an atomic assignment.
Then, there is a unique conditional assignment $\overline{h}$ extending $h$, that is such that $\overline{h}(\theta)=h(\theta)$ for any $\theta\in\Theta$\,.
\end{proposition}
The construction of $\overline{h}$ is obvious.
\paragraph{Semantic.}
Let $\mathbf{M}$ be a conditional model.
\\[5pt]
Let $(\Gamma,\Delta)\in\seqset$ be a sequent.
Then, the meta-relation $\Gamma\models_{\mathbf{M}}\Delta$ is defined by:
\begin{equation}\label{Section:Model:DBL:sem:eq:2}
\Gamma\models_{\mathbf{M}}\Delta
\mbox{ if and only if }
\forall H\in \mathcal{H}[\mathbf{M}],\;
\left[\;\forall \gamma\in\{\Gamma\},\; H(\gamma)=\Omega\;\right]
\Rightarrow
 \left[\;\exists \delta\in\{\Delta\},\; H(\delta)=\Omega\;\right]\;.
\end{equation}
The relation $\Gamma\models_{\mathbf{M}}\Delta$ means that the sequent $(\Gamma,\Delta)$ is true for the model $\mathbf{M}$.
\begin{proposition}\label{DBL:prop:modelsound:1}
Assuming $\Gamma\vdash\Delta$\,, then $\Gamma\models_{\mathbf{M}}\Delta$ for any conditional model $\mathbf{M}$.
\end{proposition}
Proof is done in appendix~\ref{Apx:Proof2Sem:sect}.
\paragraph{Model construction: purpose.}
Typically, an ultimate issue is to construct a model for which the deduction system is complete, \emph{i.e.}:
\begin{center}
\mbox{Find }$\mathbf{M}$\mbox{ such that }$\Gamma\models_{\mathbf{M}}\Delta$\mbox{ implies }$\Gamma\vdash\Delta$\,.
\end{center}
This problem is not addressed in this article.
Moreover, it is not clear that conditional models are sufficient to specify the sequents.
\\[5pt]
However, the  following completeness result is proved in section~\ref{DBL:freemodel:1}:
\begin{quote}
There is a conditional model $\mathbf{M}$ for DBL$_\ast$ such that $\models_{\mathbf{M}}\phi$ implies $\vdash\phi$ for any $\phi\in\mathcal{L}$.
\end{quote}
This model construction is applied subsequently for constructing the probabilistic extension from $\mathcal{L}_C$ to $\mathcal{L}$ for DBL or DBL$_\ast$\,:
\begin{quote}
For any probability $p$ defined over $\mathcal{L}_C$\,, there is a probability $\overline{p}$ over $\mathcal{L}$ which extends $p$ and verifies $\overline{p}(\phi\wedge\psi)=\overline{p}(\phi)\overline{p}(\psi)$, for any $\phi,\psi\in\mathcal{L}$ such that $\vdash\psi\times\phi$\,. 
\end{quote}
This result proves that DBL and DBL$_\ast$ fulfill the necessary conditions of a Bayesian logical system.
\subsection{Construction of a free conditional model for DBL$_\ast$}
\label{DBL:freemodel:1}
A free conditional model for DBL$_\ast$ is constructed now.
This model, $\mathbf{M}[\Theta]$, is such that: $\models_{\mathbf{M}[\Theta]}\phi$ implies $\vdash\phi$.
It is constructed as a direct limit of partial models.
These partial models are constructed recursively, based on the iteration of $(|)$ on any propositions.
\\[5pt]
It is recalled that $\Theta$ is a finite set.
\\[5pt]
The following result about direct limits is needed.
\subsubsection{Direct limit}
\begin{proposition}\label{DBL:proj:lim:1}
Let $K\ge1$ and let $r_k\in\Nset$ be defined for $1\le k\le K$\,.
Let be given a predicate $\varphi(E,R_1,\dots,R_K)$ defined for any set $E$ and subsets $R_k\subset E^{r_k}$.
For any $n\in\Nset$, let be defined $(E_n,R_{1,n},\dots,R_{K,n})$ and a mapping $\mu_n:E_n\rightarrow E_{n+1}$ verifying:
\begin{itemize}
\item $R_{k,n}\subset E_n^{r_k}$ for $1\le k\le K$,
\item $\mu_n:E_n\rightarrow E_{n+1}$ is one-to-one,
\item $\mu_n(R_{k,n})\subset R_{k,n+1}$\,,
\item $\varphi(E_n,R_{1,n},\dots,R_{K,n})$ holds true,
\item There is $m\ge n$ (for any $n\in\Nset$) such that $\mu_{m-1}\circ\cdots\circ\mu_n(E_n^{r_K})\subset R_{K,m}$\,.
\end{itemize}
Then there exists $(E_\infty,R_{1,\infty},\dots,R_{K,\infty})$ and a mapping sequence $\nu_n:E_n\rightarrow E_\infty$ for $n\in\Nset$, such that:
\begin{itemize}
\item $\nu_n$ is one-to-one,
\item $\nu_n=\nu_{n+1}\circ\mu_n$,
\item $\nu_n(R_{k,n})\subset R_{k,\infty}$\,, for any $n\in\Nset$,
\item $\varphi(E_\infty,R_{1,\infty},\dots,R_{K,\infty})$ holds true,
\item For any $x\in E_\infty$\,, there is $n\in\Nset$ and $y\in E_n$ such that $\nu_n(y)=x$\,,
\item $R_{K,\infty}=E_\infty^{r_K}$\,.
\end{itemize}
\end{proposition}
\begin{description}
\item[Proof.]
Classical results on \emph{direct limit} just give the property, excepted for the relation $R_{K,\infty}=E_\infty^{r_K}$\,.
\\
Now, let $(x_1,\dots,x_{r_K})\in E_\infty^{r_K}$.
\\
For $1\le k\le r_K$, there is $n_k\in\Nset$ and $y_k\in E_{n_k}$ such that $\nu_{n_k}(y_k)=x_k$\,.
\\
Let $N\in\Nset$ be such that $N\ge n_k$ for $1\le k\le r_K$\,.
\\
Then, set $z_k=\mu_{N-1}\circ\dots\circ\mu_{n_k}(y_k)$\,.
\\
It comes $(x_1,\dots,x_{r_K})=\nu_{N}(z_1,\dots,z_{r_K})$\,.
\\
Now, there is $M\ge N$ such that $\mu_{M-1}\circ\cdots\circ\mu_N(E_N^{r_K})\subset R_{K,M}\,.$
\\
Then, it is deduced $(x_1,\dots,x_{r_K})\in\nu_M(R_{K,M})\subset R_{K,\infty}$\,.
\\[5pt]
This last result just proves that $E_\infty^{r_K}\subset R_{K,\infty}$\,.
\item[$\Box\Box\Box$]\rien
\end{description}
It is noticed that operators and relations over a set $E$ could both be modeled by their graph, that is a subset of a power product of $E$.
Then, proposition~\ref{DBL:proj:lim:1} is quite general.
In particular, it makes possible the construction of structures with operators and relations as a limit of partially constructed structures.
The following corollary is implied.
\\[5pt]
{\bf Corollary.}
\emph{Let be defined a sequence $(E_n,\ast_1,\dots,\ast_K,\circ)_{n\in\Nset}$ of algebraic structures, with common algebraic properties, where the operator $\circ$ is defined on subdomains $D_n\subset E_n^r$.
Let $\mu_n:E_n\rightarrow E_{n+1}$ be a one-to-one morphism;
in particular:
\begin{itemize}
\item $\mu_n(D_n)\subset D_{n+1}$\,,
\item $\forall (x_1,\dots,x_r)\in D_n,\, \mu_n(x_1\circ\cdots\circ x_r)=\mu_n(x_1)\circ\cdots\circ \mu_n(x_r)$\,.
\end{itemize}
Assume moreover that:
\begin{equation}\label{eq:hyp:proj:1}
\mbox{there is }m\ge n\mbox{ such that }\mu_{m-1}\circ\cdots\circ\mu_n(E_n^r)\subset D_{m}\mbox{ for any }n\in\Nset\,.
\end{equation}
Then there is an algebraic structure $(E_\infty,\ast_1,\dots,\ast_n,\circ)$, where $\circ$ is entirely constructed, and one-to-one morphisms $\nu_n:E_n\rightarrow E_\infty$ such that:
\begin{itemize}
\item $\nu_n=\nu_{n+1}\circ\mu_n$,
\item For any $x\in E_\infty$\,, there is $n\in\Nset$ and $y\in E_n$ such that $\nu_n(y)=x$\,,
\item The structure $(E_\infty,\ast_1,\dots,\ast_n,\circ)$ has the same algebraic properties than the algebras of the sequence.
\end{itemize}}
\begin{description}
\item[Proof.]
It is obtained by applying the proposition to the sequence $(E_n,R_{1,n},\dots,R_{K+2,n})$ and predicate $\varphi$, where:
\begin{itemize}
\item $R_{1,n},\dots,R_{K+1,n}$ are the graphs of the operators $\ast_1,\dots,\ast_K,\circ$\,,
\item $R_{K+2,n}=D_n$\,,
\item $\varphi(E,R_{1},\dots,R_{K+2})$ encapsulates both the algebraic properties of the algebras, the functional nature of the graphs, and the domain of definition of $\circ$.
\end{itemize}
\item[$\Box\Box\Box$]\rien
\end{description}
This corollary is used now for the construction of a model of DBL$_\ast$\,.
\subsubsection{Definition of partial models}
In this section are constructed a sequence $(\mathbf{B}_n,\cup,\cap,\sim,\emptyset,\Omega_n, f_n, b_n, r_n)_{n\in\Nset}$ and a sequence of one-to-one morphisms $(\mu_n)_{n\in\Nset}$ such that:
\begin{itemize}
\item $(\mathbf{B}_n,\cup,\cap,\sim,\emptyset,\Omega_n)$ is a Boolean algebra,
\item $(\mathbf{B}_n,\cup,\cap,\sim,\emptyset,\Omega_n, f_n)$ is a partial Bayesian model; in particular, $f_n$ is partially constructed,
\item $\mu_n:\mathbf{B}_n\rightarrow \mathbf{B}_{n+1}$ is a one-to-one morphism of Bayesian models,
\item $b_n$ is an element of $\mathbf{B}_n$\,,
\item At step $n+1$, the definition of $f_{n+1}$ is completed, so as to include the domain $\mu_n(\mathbf{B}_n)\times\{\mu_n(b_n),\mu_n(\sim b_n)\}$\,,
\item $r_n:\mathbf{B}_n\rightarrow\Nset$ is a ranking function; owing to the one-to-one morphism, $r(A)$ indicates the step of construction of $A$.
\end{itemize}
The propositions $b_n$ are chosen in order to make the sequence complient with proposition~\ref{DBL:proj:lim:1} (more precisely, hypothesis~(\ref{eq:hyp:proj:1}) of the corollary).
The choice criterion is computed from the ranking function.
\\[5pt]
Then, a Bayesian model is deduced by the direct limit.
\paragraph{Notations and definitions.}
%
For any $m>n$ and $A\in \mathbf{B}_n$\,, it is defined $A_{[m}=\mu_{m-1}\circ\dots\circ\mu_{n}(A)$\,.
In the case of a subscripted propositions, say $A_k$, the notation $A_{k[m}=\bigl(A_k\bigr)_{[m}$ is used.\\[5pt]
Subsequently, a singleton $\{\omega\}$ may be denoted $\omega$ if the context is not ambiguous.
In particular, the use of the notation $\omega_{[n}$ instead of $\{\omega\}_{[n}$ is systematic.\\[5pt]
The Cartesian product of sets $A$ and $B$ is denoted $A\times B$\,;
the functions $\mathrm{id}$ and $T$ are defined over pairs by $\mathrm{id}(x,y)=(x,y)$ and $T(x,y)=(y,x)$\,;
for a set of pairs $C$, the abbreviation $(\mathrm{id}\cup T)(C)=\mathrm{id}(C)\cup T(C)$ is also used.
\paragraph{Initialization.}
Define $(\mathbf{B}_0,\cup,\cap,\sim,\emptyset,\Omega_0, f_0, b_0, r_0)$ by:
\begin{itemize}
\item $\Omega_0=\{0,1\}^{\Theta}$,
\item $\mathbf{B}_0=\mathcal{P}(\Omega_0)$ (\emph{i.e.} the set of subsets of $\Omega_0$),
\item Take $\cup,\cap,\emptyset$ as the set union, set intersection and empty set;
define $\sim$ as the set complement, that is $\sim A=\Omega_0\setminus A$\,,
\item Define $f_0(A,\emptyset)=f_0(A,\Omega_0)=A$ for any $A\in \mathbf{B}_0$\,,
\item Define $r_0(A)=0$ for any $A\in \mathbf{B}_0$\,,
\item Choose $b_0\in\mathbf{B}_0\setminus\{\emptyset,\Omega_0\}$\,.
{\bf It is noticed that $b_{0}\not\in\{\emptyset,\Omega_{0}\}$.}
\end{itemize}
\paragraph{Step $n$ to step $n+1$.}
Let $(\mathbf{B}_k,\cup,\cap,\sim,\emptyset,\Omega_k, f_k, b_k, r_k)_{0\le k\le n}$ and the one-to-one morphisms $(\mu_k)_{0\le k\le n-1}$ be constructed.
\\[5pt]
Then, construct the set $I_n$ and the sequences $\Gamma_n(i),\Pi_n(i)|_{i\in I_n}$ according to the cases:
\subparagraph{Case 0.} There is $m<n$ such that $\{b_{m[n},\sim b_{m[n}\}=\{b_n,\sim b_n\}$\,.\\
Let $\nu$ be the greatest of such $m$.
{\bf Notice that the hypothesis $b_n=b_{\nu[n}$ holds by construction~(\ref{bn:construc:eq:1})}.
Then define $I_n=\mu_\nu(b_{\nu})\times\sim \mu_\nu(b_{\nu})$\,,\\
$\Pi_n(\omega,\omega')=f_n(\omega'_{[n},\sim b_{n})\cap\omega_{[n}$ and $\Gamma_n(\omega,\omega')=f_n(\omega_{[n},b_{n})\cap\omega'_{[n}$ for any $(\omega,\omega')\in I_n$\,.
\\[5pt]
Remark: case 0 means that the construction of $f(\cdot,b_n)$ and of $f(\cdot,\sim b_n)$ has already begun over the propositions of $\mathbf{B}_{\nu+1}$.
\subparagraph{Case 1.} Case 0 does not hold;\\
Define $I_n=\{b_n\}$\,, $\Pi_n(i)=i$ 
and $\Gamma_n(i)=\sim i$ for any $i\in I_n$\,.
\\[5pt]
Remark: case 1 means that $f(\cdot,b_n)$ and $f(\cdot,\sim b_n)$ are constructed for the first time.
\subparagraph{Setting.}
$(\mathbf{B}_{n+1},\cup,\cap,\sim,\emptyset,\Omega_{n+1}, f_{n+1}, b_{n+1}, r_{n+1})$ and $\mu_n$ are defined by:
\begin{itemize}
\item $\mu_n(A)=\bigcup_{i\in I_n} \biggl(\Bigl(\bigl(A\cap\Pi_n(i)\bigr)\times\Gamma_n(i)\Bigr)
\cup \Bigl(\bigl(A\cap\Gamma_n(i)\bigr)\times\Pi_n(i)\Bigr)\biggr)$ for any $A\in \mathbf{B}_{n}$\,,
\item $\Omega_{n+1}=\mu_n(\Omega_n)$\,,
\item $\mathbf{B}_{n+1}=\mathcal{P}(\Omega_{n+1})$\,,
\item Take $\cup,\cap,\emptyset$ as the set union, set intersection and empty set;
define $\sim$ as the set complement, that is $\sim A=\Omega_{n+1}\setminus A$\,,
\item $f_{n+1}(A,\emptyset)=f_{n+1}(A,\Omega_{n+1})=A$ for any $A\in \mathbf{B}_{n+1}$\,,
\item For any $A\in \mathbf{B}_n\setminus\{b_n,\sim b_n,\emptyset,\Omega_n\}$ and any $B\in \mathbf{B}_n$ such that $f_n(B,A)$ is defined, then $f_{n+1}\bigl(\mu_n(B),\mu_n(A)\bigr)$ is defined and $f_{n+1}\bigl(\mu_n(B),\mu_n(A)\bigr)=\mu_n\bigl(f_n(B,A)\bigr)$\,,
\item For any $A\in \mathbf{B}_{n+1}$\,, set
$
f_{n+1}\bigl(A,\mu_n(b_n)\bigr)=(\mathrm{id}\cup T)
\biggl(A\cap\Bigl(\bigcup_{i\in I_n}\bigl(\Pi_n(i)\times\Gamma_n(i)\bigr)\Bigr)\biggr)
$\\[5pt]
and
$
f_{n+1}\bigl(A,\sim\mu_n(b_n)\bigr)=(\mathrm{id}\cup T)
\biggl(A\cap\Bigl(\bigcup_{i\in I_n}\bigl(\Gamma_n(i)\times\Pi_n(i)\bigr)\Bigr)\biggr)
\;,
$
\item Define $r_{n+1}(\mu_n(A))=r_n(A)$ for any $A\in \mathbf{B}_n$ and $r_{n+1}(A)=n+1$ for any $A\in\mathbf{B}_{n+1}\setminus\mu_n(\mathbf{B}_n)$\,,
($r_{n+1}$ just maps to the first step of occurence of the proposition)
\item Define:
\begin{equation}
\label{bn:construc:eq:0:1}\begin{array}{@{}l@{}}\displaystyle\vspace{4pt}
\widetilde{b}_{n+1}\in\arg\min_{B\in\mathbf{B}_{n+1}}\lambda_{n+1}(B)\;,\mbox{ where:}
\\\rien\hspace{20pt}\displaystyle
\lambda_{n+1}(B)=\inf\left\{
r_{n+1}(A)+r_{n+1}(B)
\;\left/\;A\in\mathbf{B}_{n+1}\mbox{ and }f(A,B)\mbox{ is undefined}
\right.\right\}\;.
\end{array}\end{equation}
Then, define $b_{n+1}$ by:
\begin{equation}\label{bn:construc:eq:1}
\begin{array}{@{}l@{}}\displaystyle
b_{n+1}=b_{m[n+1} \mbox{ if there is }m\le n\mbox{ such that }\bigl\{\widetilde{b}_{n+1},\sim \widetilde{b}_{n+1}\bigr\}=\bigl\{b_{m[n+1},\sim b_{m[n+1}\bigr\}\;,
\\\displaystyle
b_{n+1}=\widetilde{b}_{n+1} \mbox{ otherwise}.
\end{array}\end{equation}
The purpose of equation~(\ref{bn:construc:eq:0:1}) is to choose $b_{n+1}$ (or its negation) in order to continue the construction of $f$ on the oldest pairs first.
By doing that, the condition~(\ref{eq:hyp:proj:1}) of the direct limit is ensured.
The purpose of equation~(\ref{bn:construc:eq:1}) is to choose $b_{n+1}$ in coherence with a possible previous occurence.
{\bf It is noticed that $b_{n+1}\not\in\{\emptyset,\Omega_{n+1}\}$.}
\end{itemize}
\paragraph{Short explanation of the model.}
In fact, $(\omega,\omega')\in\Pi_n(i)\times\Gamma_n(i)$ should be interpreted as $\omega\wedge(\omega'|\neg b_n)$, while $(\omega',\omega)\in\Gamma_n(i)\times\Pi_n(i)$ should be interpreted as $\omega'\wedge(\omega|b_n)$.
The reader should compare this construction to the proof of completeness in appendix~\ref{Appendix:ProofOfAlmostCompletude} for a better comprehension of the mechanisms of the model.
\subsubsection{Properties of $(\mathbf{B}_{n},\cup,\cap,\sim,\emptyset,\Omega_{n}, f_{n}, \mu_n)_{n\in\Nset}$}
\label{Omega:properties}
It is proved recursively:
\begin{description}
\item[$\rien\quad\alpha1$.] $\mu_n:\mathbf{B}_n\rightarrow \mathbf{B}_{n+1}$ is a one-to-one Boolean morphism,
\item[$\rien\quad\alpha2$.] If $A,B\in \mathbf{B}_n$ and $f_n(B,A)$ is defined, then $f_{n+1}\bigl(\mu_n(B),\mu_n(A)\bigr)=\mu_n\bigl(f_n(B,A)\bigr)$\,,
\item[$\rien\quad\beta1$.] Let $A,B\in \mathbf{B}_n$ such that $f_n(B,A)$ is defined.
\\Then $A\subset B$ and $A\ne\emptyset$ imply $f_n(B,A)=\Omega_n$\,,
\item[$\rien\quad\beta2$.] Let $A,B,C\in \mathbf{B}_n$ such that $f_n(B,A)$, $f_n(C,A)$ and $f_n(B\cup C,A)$ are defined.
\\Then $f_n(B\cup C,A)= f_n(B,A)\cup f_n(C,A)$\,,
\item[$\rien\quad\beta3$.] Let $A,B\in \mathbf{B}_n$ such that $f_n(B,A)$ is defined.
\\Then $A\cap f_n(B,A) = A\cap B$\,,
\item[$\rien\quad\beta4$.] Let $A,B\in \mathbf{B}_n$ such that $f_n(B,A)$ and $f_n(\sim B,A)$ are defined.
\\Then $f_n(\sim B,A)=\sim f_n(B,A)$\,,
\item[$\rien\quad\beta5w$.] Let $A,B\in \mathbf{B}_n$ such that $f_n(B,A)$ and $f_n(B,\sim A)$ are defined.
\\Then $f_n(B,A)=B$ implies $f_n(B,\sim A)=B$\,.
\end{description}
\emph{Proofs are given in appendix~\ref{Appendix:MainProof}.}
\subsubsection{Limit}
Corollary of proposition~\ref{DBL:proj:lim:1} applies to the sequence $(\mathbf{B}_{n},\cup,\cap,\sim,\emptyset,\Omega_{n}, f_{n}, \mu_n)_{n\in\Nset}$ \,.
In particular, the condition~(\ref{eq:hyp:proj:1}) is derived from: $$\lim_{n\rightarrow+\infty}\;\;\min_{B\in\mathbf{B}_n}\lambda_n(B)=+\infty\;,$$
which itself is a consequence of~(\ref{bn:construc:eq:0:1}) and the construction process.
\\[5pt]
As a consequence, there is a Bayesian model $\mathbf{M}[\Theta]=(\mathbf{B}[\Theta],\cup,\cap,\sim,\emptyset,\Omega, f)$ and a sequence $(\nu_n)_{n\in\Nset}$ such that:
\begin{itemize}
\item $\nu_n:\mathbf{B}_n\rightarrow\mathbf{B}[\Theta]$ is a one-to-one morphism of Bayesian model,
\item $\nu_n=\nu_{n+1}\circ\mu_n$,
\item For any $A\in \mathbf{B}[\Theta]$\,, there is $n\in\Nset$ and $A_n\in \mathbf{B}_{n}$ such that $\nu_n(A_n)=A$\,.
\end{itemize}
\subsubsection{Completeness for the conditional operator}
For any $\theta\in\Theta$\,, define $\xi_\theta\in\mathbf{B}_0$ by $\xi_\theta=\bigl\{(\delta_\tau)_{\tau\in\Theta}\in\Omega_0\,\big/\,\delta_\theta=1\bigr\}$\,.
Then, define the atomic assignment $h:\Theta\rightarrow\mathbf{B}[\Theta]$ by $h(\theta)=\nu_0(\xi_\theta)$ for any $\theta\in\Theta$\,.
Denote $\overline{h}$ the extention of $h$ toward $\mathcal{L}$.
\begin{proposition}\label{DBL:Complet:prop:1}
Let $\phi\in\mathcal{L}_C$\,.
Then, $\vdash_C\phi$ if and only if $\overline{h}(\phi)=\Omega$\,.
\end{proposition}
Proof is obvious since $\mathbf{B}_0$ is isomorph to the factor set of $\mathcal{L}_C$ with respect to $\equiv_C$.
\begin{proposition}\label{DBL:Complet:prop:2}
Let $\phi\in\mathcal{L}$\,.
Then, the following assertions are equivalent:
\begin{itemize}
\item $\vdash\phi$ in DBL$_\ast$\,,
\item $\overline{h}(\phi)=\Omega$\,,
\item $\models_{\mathbf{M}[\Theta]}\phi$\,.
\end{itemize}
\end{proposition}
Proof is done in appendix~\ref{Appendix:ProofOfAlmostCompletude}\,.
\\[5pt]
Proposition 2 expresses that $\mathbf{M}[\Theta]$ is complete for the conditional operator.
\begin{proposition}\label{DBL:Complet:prop:3}
Let $\phi\in\mathcal{L}_C$\,, such that $\vdash\phi$ in DBL$_\ast$\,.
Then $\vdash_C\phi$\,.
\end{proposition}
Obvious from~\ref{DBL:Complet:prop:1} and~\ref{DBL:Complet:prop:2}.
\\[5pt]
This result proves that $DBL_\ast$ is an extension of classical logic.
Now, proposition~\ref{DBL:Complet:prop:2} implies that $DBL_\ast$ is much more than just classical logic:
\begin{proposition}\label{DBL:Complet:prop:4}
\emph{[Non-distortion property]}
Let $\phi,\psi\in\mathcal{L}_C$.
Assume that $\vdash\phi,\psi$ in DBL$_\ast$.
Then $\vdash_C\phi$ or $\vdash_C\psi$.
\end{proposition}
Interpretation: DBL$_\ast$ does not ``distort'' the \emph{classical} propositions.
More precisely, a property like $\vdash\phi,\psi$ would add some knowledge about $\phi$ and $\psi$.
But the \emph{non-distortion} just tells that it is impossible unless there is a trivial knowledge about $\phi$ or $\psi$ within classical logic.
\begin{description}
\item[Proof.]
Assume $\vdash\phi,\psi$\,.\\
Since $\mathbf{M}[\Theta]$ is a model for DBL$_\ast$, it comes that $H(\phi)=\Omega$ or $H(\psi)=\Omega$ for any $H\in\mathcal{H}\bigl[\mathbf{M}[\Theta]\bigr]$\,.\\
In particular, $\overline{h}(\phi)=\Omega$ or $\overline{h}(\psi)=\Omega$ (definition~(\ref{Section:Model:DBL:sem:eq:2})\,)\,.
\\
Since $\phi\in\mathcal{L}_C$ and $\psi\in\mathcal{L}_C$, it comes $\vdash_C\phi$ or $\vdash_C\psi$\,.
\item[$\Box\Box\Box$]\rien
\end{description}
Another \emph{non-distortion} property is derived subsequently in the context of probabilistic DBL$_\ast$\,.
\section{Extension of probability}
\label{Section:Proba:DBL}
\subsection{Probability over propositions}
%
Probabilities are classically defined over measurable sets.
However, this is only a manner to model the notion of probability, which is essentially an additive measure of the belief of logical propositions.
Probability could be defined without reference to the measure theory, at least when the propositions are countable.
The notion of probability is explained now within a strict propositional formalism.
Conditional probabilities are excluded from this definition, but the notion of independence is considered.
\vspace{5pt}\\
Intuitively, a probability over a space of logical propositions is a measure of belief which is additive (disjoint propositions are adding their chances) and increasing with the propositions.
This measure should be zeroed for the contradiction and set to $1$ for the tautology.
Moreover, \emph{a probability is a multiplicative measure for independent propositions}.
%
\paragraph{Definition for classical propositions.}
A probability $\pi$ over $C$\,, the classical logic, is a $\Rset^+$ valued function such that for any propositions $\phi$ and $\psi$ of $\mathcal{L}_C$\,:
\begin{description}
\item[\rien$\quad$\emph{Equivalence.}]$\phi\equiv_C\psi$ implies $\pi(\phi)=\pi(\psi)$\,,
\item[\rien$\quad$\emph{Additivity.}]$\pi(\phi\wedge\psi)+\pi(\phi\vee\psi)=\pi(\phi)+\pi(\psi)$\,,
\item[\rien$\quad$\emph{Coherence.}]$\pi(\bot)=0$\,,
\item[\rien$\quad$\emph{Finiteness.}]$\pi(\top)=1$\,.
\end{description}
\subparagraph{Property.}
The coherence and additivity implies the increase of $\pi$:
\begin{description}
\item[\rien$\quad$\emph{Increase.}]$\pi(\phi\wedge\psi)\le \pi(\phi)$\,.
\end{description}
\begin{description}
\item[Proof.] 
Since $\phi\equiv_C(\phi\wedge\psi)\vee(\phi\wedge\neg\psi)$ and $(\phi\wedge\psi)\wedge(\phi\wedge\neg\psi)\equiv_C\bot$, the additivity implies:
$$
\pi(\phi)+\pi(\bot)=\pi(\phi\wedge\psi)+\pi(\phi\wedge\neg\psi)\;.
$$
From the coherence $\pi(\bot)=0$\,,
it is deduced $\pi(\phi)=\pi(\phi\wedge\psi)+\pi(\phi\wedge\neg\psi)$\,.\\
Since $\pi$ is non-negatively valued, $\pi(\phi)\ge \pi(\phi\wedge\psi)$\,.
\item[$\Box\Box\Box$]\rien
\end{description}
\paragraph{Definition for DBL/DBL$_\ast$.}
In this case, we have to deal with independence notions.\\[5pt]
A probability $P$ over DBL/DBL$_\ast$ is a $\Rset^+$ valued function, which verifies (replace $\equiv_C$ by $\equiv$ and $\pi$ by $P$) \emph{equivalence}, \emph{additivity}, \emph{coherence}, \emph{finiteness} and:
\begin{description}
\item[\rien$\quad$\emph{Multiplicativity.}]$\vdash\phi\times\psi$ implies $P(\phi\wedge\psi)=P(\phi)P(\psi)$\,.
\end{description}
for any propositions $\phi$ and $\psi$ of $\mathcal{L}$\,. 
\subsection{Probability extension over DBL$_\ast$}
\label{ProbExt:wDBL}
\begin{proposition}\label{dbl:prob:ext:1}
Let $\pi$ be a probability defined over $C$\,, the classical logic, such that $\pi(\phi)>0$ for any $\phi\not\equiv_C\bot$.
Then, there is a (multiplicative) probability $\overline{\pi}$ defined over DBL$_\ast$ such that $\overline{\pi}(\phi)=\pi(\phi)$ for any classical proposition $\phi\in\mathcal{L}_C$\,.
\end{proposition}
\emph{Remark: this is another non-distortion property, since the construction of DBL$_\ast$ puts no constraint over probabilistic classical propositions.}
\\[5pt]
Proof is done in appendix~\ref{Appendix:Probabilition}.
\\[5pt]
{\bf Corollary.}
Let $\pi$ be a probability defined over $C$\,.
Then, there is a (multiplicative) probability $\overline{\pi}$ defined over DBL$_\ast$ such that $\overline{\pi}(\phi)=\pi(\phi)$ for any $\phi\in\mathcal{L}_C$\,.
\begin{description}
\item[Proof.]
Let $\Sigma=\left\{\left.\bigwedge_{\theta\in\Theta}\epsilon_\theta\;\right/\;\epsilon\in\prod_{\theta\in\Theta}\{\theta,\neg\theta\}\right\}$\,, a generating partition of $\mathcal{L}_C$.\\
For any real number $e >0$\,, define the probability $\pi_e $ over $\mathcal{L}_C$ by:
$$
\forall\sigma\in\Sigma\,,\; \pi_e (\sigma)=\frac{e }{\mathrm{card}(\Sigma)}+(1-e )\pi(\sigma)
\;. 
$$
Let $\overline{\pi_e}$ be the extension of  $\pi_e$ over DBL$_\ast$ as constructed in appendix~\ref{Appendix:Probabilition}.\\
It is noticed in~\ref{AppC:Conclude}\,, that there is by construction a rational function $R_\phi$ such that $\overline{\pi_e}(\phi)=R_\phi(e )$ for any $\phi\in\mathcal{L}$\,.\\
Now $0\le R_\phi(e )\le 1$\,;
since $R_\phi(e )$ is rational and bounded, $\lim_{e \rightarrow 0+}R_\phi(e )$ exists.\\
Define $\overline{\pi}(\phi)=\lim_{e \rightarrow 0+}R_\phi(e )$\,, for any $\phi\in\mathcal{L}$.\\
The additivity, coherence, finiteness and multiplicativity are then inherited by $\overline{\pi}$.\\
At last, it is clear that $\overline{\pi}(\sigma)=\pi(\sigma)$ for any $\sigma\in\Sigma$\,.
\item[$\Box\Box\Box$]\rien
\end{description}
\subsection{Model and probability extension for DBL}
Let $\mathcal{K}$ be the set of all (multiplicative) probabilities $P$ over DBL$_\ast$ such that $P(\phi)>0$ for any $\phi\not\equiv\bot$\,,
and define the sequences $\mathcal{K}(\phi)=(P(\phi))_{P\in\mathcal{K}}$ for any $\phi\in\mathcal{L}$\,.\\
Then define
$\mathcal{L}_{\mathcal{K}}=\mathcal{K}(\mathcal{L})=\bigl\{\mathcal{K}(\phi)\;\big/\;\phi\in\mathcal{L}\bigr\}\;;$
The space $\mathcal{L}_{\mathcal{K}}$ is thus a subset of ${\Rset^+}^{\mathcal{K}}$\,.\\
The operators $\neg$, $\wedge$, $\vee$ and $(|)$ are canonically implied over $\mathcal{L}_{\mathcal{K}}$\,:
$$\begin{array}{@{}l@{}}\displaystyle
\neg\mathcal{K}(\phi)=\mathcal{K}(\neg\phi)\,,\ 
\mathcal{K}(\phi)\wedge\mathcal{K}(\psi)=\mathcal{K}(\phi\wedge\psi)\,,\ 
\mathcal{K}(\phi)\vee\mathcal{K}(\psi)=\mathcal{K}(\phi\vee\psi)\hspace{50pt}\rien
\gonextline\displaystyle
\mbox{and}\quad
\bigl(\mathcal{K}(\psi)\big|\mathcal{K}(\phi)\bigr)=
\mathcal{K}\bigl((\psi|\phi)\bigr)\;.
\end{array}$$
Since any $P\in\mathcal{K}$ verifies the equivalence property, it comes $\mathcal{K}(\phi)=\mathcal{K}(\psi)$ when $\phi\equiv\psi$ in DBL$_\ast$.
In particular, $\bigl(\mathcal{L}_{\mathcal{K}},\vee,\wedge,\neg,\mathcal{K}(\bot),\mathcal{K}(\top))$ is a Boolean algebra.
\begin{proposition}\label{dbl:prob:ext:2}
$\bigl(\mathcal{L}_{\mathcal{K}},\vee,\wedge,\neg,\mathcal{K}(\bot),\mathcal{K}(\top),(|)\bigr)$ is a conditional model of DBL.
\end{proposition}
\begin{description}
\item[Proof.]
Since $\phi\equiv\psi$ in DBL$_\ast$ implies $\mathcal{K}(\phi)=\mathcal{K}(\psi)$\,, it comes $\beta2$, $\beta3$ and $\beta4$, from axioms b2, b3 and b4.
\item[\emph{Proof of $\beta1$.}]
Assume $\mathcal{K}(\phi\rightarrow\psi)=\mathcal{K}(\top)$ and $\mathcal{K}(\phi)\ne\mathcal{K}(\bot)$\,.
\\[5pt]
Let $P\in\mathcal{K}$\,.\\
Then $P(\neg\phi\vee\psi)=P(\phi\rightarrow\psi)=P(\top)=1$\,.\\
Now $P(\neg\phi\vee\psi)=1$ implies $P(\phi\wedge\psi)+P(\neg\phi)=1$\,.\\
As a consequence, $P(\phi\wedge\psi)=1-P(\neg\phi)=P(\phi)$.\\
Now, Hypothesis $\mathcal{K}(\phi)\ne\mathcal{K}(\bot)$ implies $\phi\not\equiv\bot$ and then $P(\phi)\ne0$\,.\\
Since $P$ is multiplicative and $P(\phi)\ne0$\,, it comes $P\bigl((\psi|\phi)\bigr)=P(\phi\wedge\psi)/P(\phi)=1$\,.
\\[5pt]
At last, $\mathcal{K}(\psi|\phi)=\mathcal{K}(\top)$ and, consequently, $\bigl(\mathcal{K}(\psi)\big|\mathcal{K}(\phi)\bigr)=\mathcal{K}(\top)$\,.\\
The model verifies $\beta1$.
\item[\emph{Proof of $\beta5$.}]
Since $P$ is multiplicative for any $P\in\mathcal{K}$\,, $\vdash(\psi|\phi)\times\phi$ and $(\psi|\phi)\wedge\phi\equiv\psi\wedge\phi$ in DBL$_\ast$, it comes $P\bigl((\psi|\phi)\bigr)P(\phi)=P(\psi\wedge\phi)\mbox{ for any }P\in\mathcal{K}\,.$
\\
Now assume $\bigl(\mathcal{K}(\psi)\big|\mathcal{K}(\phi)\bigr)=\mathcal{K}(\psi)$\,, with $\psi\not\equiv\bot$\,.\\
Then $\mathcal{K}\bigl((\psi|\phi)\bigr)=\mathcal{K}(\psi)$, and $P\bigl((\psi|\phi)\bigr)=P(\psi)$ for any $P\in\mathcal{K}$\,.\\
Then $P(\phi)=P(\psi\wedge\phi)/P\bigl((\psi|\phi)\bigr)=P(\psi\wedge\phi)/P(\psi)=P\bigl((\phi|\psi)\bigr)$ for any $P\in\mathcal{K}$\,,
\\
and $\bigl(\mathcal{K}(\phi)\big|\mathcal{K}(\psi)\bigr)=\mathcal{K}\bigl((\phi|\psi)\bigr)=\mathcal{K}(\phi)$\,.\\
Since moreover $(\phi|\bot)\equiv\phi$ and $(\bot|\phi)\equiv\bot$ in DBL$_\ast$, the model verifies $\beta5$\,.
\item[$\Box\Box\Box$]\rien
\end{description}
{\bf Corollary.}
$\phi\equiv\psi$ in DBL implies $\mathcal{K}(\phi)=\mathcal{K}(\psi)$.
\paragraph{Probability extension.}
The corollary implies that any $P\in\mathcal{K}$ is a (multiplicative) probability over DBL.
Now, the probability extensions constructed in appendix~\ref{Appendix:Probabilition} are also elements of $\mathcal{K}$\,.
As a consequence, the proposition~\ref{dbl:prob:ext:1} as well as its corollary  still work in DBL:
\begin{proposition}\label{dbl:dbl:prob:ext:1}
Let $\pi$ be a probability defined over $C$\,, the classical logic.
Then, there is a (multiplicative) probability $\overline{\pi}$ defined over DBL such that $\overline{\pi}(\phi)=\pi(\phi)$ for any $\phi\in\mathcal{L}_C$\,.
\end{proposition}
\paragraph{Non-distortion.}
\begin{proposition}\label{dbl:dbl:prob:nondist:2}
Let $\phi,\psi$ be classical propositions.
Assume that $\vdash\phi,\psi$ in DBL.
Then $\vdash_C\phi$ or $\vdash_C\psi$.
\end{proposition}
\begin{description}
\item[Proof.]
Notice first that $\mathcal{K}:\phi\mapsto\mathcal{K}(\phi)$ is a conditional assignment by construction.
\\
From proposition~\ref{dbl:prob:ext:2}, $\vdash\phi,\psi$ implies $\mathcal{K}(\phi)=\mathcal{K}(\top)$ or $\mathcal{K}(\psi)=\mathcal{K}(\top)$\,.
\\
It follows $\forall P\in\mathcal{K}\,,\; P(\phi)=1$ or $\forall P\in\mathcal{K}\,,\; P(\psi)=1$\,.
\\
By the probability extension:
$\forall \pi\,,\; \pi(\phi)=1$ or $\forall \pi\,,\; \pi(\psi)=1$\,, where $\pi$ denotes any probability over $C$\,.\\
At last, $\vdash_C\phi$ or $\vdash_C\psi$\,.
\item[$\Box\Box\Box$]\rien
\end{description}
\subsection{Properties of the conditional}
\paragraph{Bayes inference.}
Assume a (multiplicative) probability $P$ defined over DBL/DBL$_\ast$.
Define $P(\psi|\phi)$ as an abbreviation for $P\bigl((\psi|\phi)\bigr)$\,.
Then:
$$
P(\psi|\phi)P(\phi)=P(\phi\wedge\psi)\;.
$$
\begin{description}
\item[Proof.]A consequence of \mbox{$(\psi|\phi)\wedge\phi\equiv\phi\wedge\psi$} and \mbox{$\vdash(\psi|\phi)\times\phi$}\,.
\item[$\Box\Box\Box$]\rien
\end{description}
\paragraph{About Lewis' triviality.}
The previous extension theorems has shown that for any probability $\pi$ defined over $C$\,, it is possible to construct a (multiplicative) probability $\overline{\pi}$ over DBL which extends $\pi$.
This result by itself shows that DBL avoids Lewis' triviality.
But a deeper explanation seems necessary.
\\[5pt]
Assume $\phi\in\mathcal{L}_C$ and define the probability $\pi_\phi$ over $C$ by $\pi_\phi=\pi(\cdot|\phi)$\,.
Let $\overline{\pi_\phi}$ be the extension of $\pi_\phi$ over DBL.
It happens that $\overline{\pi_\phi}\ne \overline{\pi}(\cdot|\phi)$\,, which implies that Lewis' triviality does not work anymore.
It is noticed that although $\overline{\pi}(\cdot|\phi)$ is a probability over DBL in the classical meaning (it is \emph{additive}, \emph{coherent} and \emph{finite}), it is not necessarily \emph{multiplicative}.
\begin{quote}
Conditional probabilities do not recognize the logical independence and the logical conditioning.
\end{quote}
This limitation is unavoidable: otherwise the derivation~(\ref{Eq:DBL:v2:Lewis:1}) of the triviality is possible, even if $(\psi|\phi)$ is not equivalent to a classical proposition.
\section{Conclusion}
\label{Fus2004::Sec:8}
In this contribution, the conditional logics DBL and DBL$_\ast$, a slight relaxation of DBL, have been defined and studied.
These logics have been introduced as an abstraction and extrapolation of general probabilistic properties.
DBL and DBL$_\ast$ implement the essential ingredients of the Bayesian inference, including the classical nature of the sub-universe, the inference property and a related concept of logical independence.
\\[5pt]
The logics are coherent and non-trivial.
A model has been constructed for the logic DBL$_\ast$, and completeness results have been derived.
It has been shown that any probability over the classical propositions could be extended to DBL/DBL$_\ast$, in compliance with the independence relation.
Then, the probabilistic Bayesian rule has been recovered from DBL/DBL$_\ast$.
\\[5pt]
There are still many open questions.
For example, it is possible to bring some enrichment to the conditional of DBL, by means of additional axioms.
It is also possible to consider other valuation mechanisms than the probabilities.
As a perspective, many decision systems for manipulating uncertain information could be derived from this principle.
From the strict logical viewpoint, the Deterministic Bayesian Logic also offers some interesting properties.
In particular, the notion of independence in DBL have nice logical consequences in the deductions (\emph{e.g.} regularity with an inference).
This property should be of interest in mathematical logic.
%
%
%

%
\appendix
\section{Classical subsystem}
\label{proof:clasrestric}
In this section is considered the deduction subsystem made up of rules CUT and STRUCT combined with axioms modus ponens and $c\ast$.
It is first shown that this subsystem, called classical subsystem of DBL, infers the classical tautologies.
However, the system is essentially different from the sequent calculus LK and differences are analysed.
Subsequently, the notation $\vdash_C$ is used in order to indicate that the proofs are derived within the classical subsystem of DBL.
\begin{center}
\emph{These properties also hold true within DBL, while replacing $\vdash_C$ by $\vdash$.}
\end{center}
\begin{proposition}\label{app:clarest:1}
Assuming $\vdash_C\phi$ and $\vdash_C\phi\rightarrow\psi$, it is deduced $\vdash_C\psi$\,.
\end{proposition}
This is just a modus ponens rule for sequents of the form $\vdash_C\phi$.
\begin{description}
\item[Proof.]
Obtained by applying CUT and the axiom modus ponens.
\item[$\Box\Box\Box$]\rien
\end{description}
\begin{proposition}\label{app:clarest:2}
Assume that $\phi$ is a tautology of classical logic.
Then $\vdash_C\phi$ is deduced from the classical subsystem of DBL.
\end{proposition}
\begin{description}
\item[Proof.]
It is known that tautologies of classical logic are obtained by applying the modus ponens and the axioms $\phi\rightarrow(\psi\rightarrow\phi)$\,,
$(\eta\rightarrow(\phi\rightarrow\psi))\rightarrow((\eta\rightarrow\phi)\rightarrow(\eta\rightarrow\psi))$ and $(\neg\phi\rightarrow\neg\psi)\rightarrow((\neg\phi\rightarrow\psi)\rightarrow\phi)$\,.
\\
Proposition~\ref{app:clarest:2} is then a consequence of propositin~\ref{app:clarest:1} and axioms $c1, c2, c3$.
\item[$\Box\Box\Box$]\rien
\end{description}
Proposition~\ref{app:clarest:2} has shown that the classical subsystem of DBL infers the classical tautologies, actually expressed as sequents of only one right formula.
Now, the case of general sequents is not managed by proposition~\ref{app:clarest:2}.
It happens that there are strong difference between DBL and the classical sequent calculus LK.
Although most rules of LK could be derived from the classical subsystem of DBL, some rules for manipulating the disjunction and the negation do not hold anymore.
\\[5pt]
The following table enumerates rules of LK~\cite{girard}, derived from DBL.
The table also indicates the classical tautologies, from which the rules are derived, and intermediate sequents in this derivation (several CUTs are used):
$$\begin{array}{|c|c|c|}
\hline
\mbox{LK rule}
&
\mbox{Original}
&
\mbox{Intermediate}
\\
&
\mbox{tautology}
&
\mbox{sequent}
\\\hline
\seqcalcZ{\phi\vdash\phi}\ (I)
&
\vdash_C\phi\rightarrow\phi
&
\phi\vdash_C\phi
\\\hline
\seqcalc{\Gamma,\phi\vdash\Delta}{\Gamma,\phi\wedge\psi\vdash\Delta}\ (\wedge L)
&
\vdash_C(\phi\wedge\psi)\rightarrow\phi
&
\phi\wedge\psi\vdash_C\phi
\\\hline
\seqcalc{\Gamma\vdash\phi,\Delta}{\Gamma\vdash\phi\vee\psi,\Delta}\ (\vee R)
&
\vdash_C\phi\rightarrow(\phi\vee\psi)
&
\phi\vdash_C\phi\vee\psi
\\\hline
\seqcalcT{\Gamma\vdash\phi,\Delta}{\Sigma\vdash\psi,\Pi}{\Gamma,\Sigma\vdash\phi\wedge\psi,\Delta,\Pi}\ (\wedge R)
&
\vdash_C\phi\rightarrow\bigl(\psi\rightarrow(\phi\wedge\psi)\bigr)
&
\phi,\psi\vdash_C\phi\wedge\psi
\\\hline
\seqcalcT{\Gamma\vdash\phi,\Delta}{\Sigma,\psi\vdash\Pi}{\Gamma,\Sigma,\phi\rightarrow\psi\vdash\Delta,\Pi}\ (\rightarrow L)
&
&
\phi,\phi\rightarrow\psi\vdash_C\psi\ \mathrm{(m.p.)}
\\\hline
\seqcalc{\Gamma\vdash\phi,\Delta}{\Gamma,\neg\phi\vdash\Delta}\ (\neg L)
&
\vdash_C\phi\rightarrow(\neg\phi\rightarrow\bot)
&
\phi,\neg\phi\vdash_C
\\\hline
\end{array}$$
The deduction of $\neg L$ is typical and is illustrated now.
\begin{description}
\item[Derivation of $\neg L$.]
First at all, $\vdash_C\phi\rightarrow(\neg\phi\rightarrow\bot)$ is derived as a tautology.
\\
Applying CUT and axiom m.p., it comes $\phi,\neg\phi\vdash_C\bot$\,.
\\
Applying STRUCT, it comes $\phi,\neg\phi\vdash_C$\,.
\\
Using CUT together with $\Gamma\vdash_C\phi,\Delta$\,, it is derived $\Gamma,\neg\phi\vdash_C\Delta$\,.
\item[$\Box\Box\Box$]\rien
\end{description}
On the other hand, the rules:
$$
\seqcalcT{\Gamma,\phi\vdash\Delta}{\Sigma,\psi\vdash\Pi}{\Gamma,\Sigma,\phi\vee\psi\vdash\Delta,\Pi}\ (\vee L)
\ ,\ 
\seqcalc{\Gamma,\phi\vdash\psi,\Delta}{\Gamma\vdash\phi\rightarrow\psi,\Delta}\ (\rightarrow R)
\ ,\ 
\seqcalc{\Gamma,\phi\vdash\Delta}{\Gamma\vdash\neg\phi,\Delta}\ (\neg R)
$$
of LK cannot be deduced from DBL.
In particular, the sequents:
\begin{equation}\label{Non-trueSeq:eq:1}
\phi\vee\psi\vdash\phi,\psi
\mbox{ and }\vdash\phi,\neg\phi\mbox{ are not implied by DBL (\emph{e.g.} refer to proposition~\ref{dbl:dbl:prob:nondist:2}).}
\end{equation}
These facts are obtained from the model construction of DBL.
\section{Proof: the logical theorems}
\label{proof:logth}
For concision, details of the sequent derivations are omitted, in particular concerning classical deductions.
Appendix~\ref{proof:clasrestric} establishes some important facts about the classical deduction within DBL.
It is noticed that some deductions of the sequent calculus LK are not allowed.
The rules of DBL are still powerful and sufficient though.
\paragraph{Axioms order.}
By b4, $\neg\phi\times\psi=(\neg\phi|\psi)\leftrightarrow\neg\phi\equiv\neg(\phi|\psi)\leftrightarrow\neg\phi\equiv
(\phi|\psi)\leftrightarrow\phi=\phi\times\psi$.\\
Then $\neg\phi\times\psi\vdash\phi\times\psi$ and $\phi\times\psi\vdash\neg\phi\times\psi$\,, by applying modus ponens and CUT.\\
Now, b5 implies $\psi\times\phi\vdash\phi\times\psi$, $\phi\times\psi\vdash\psi\times\phi$, $\neg\phi\times\psi\vdash\psi\times\neg\phi$ and $\psi\times\neg\phi\vdash\neg\phi\times\psi$.\\
Applying CUT then yields  $\psi\times\phi\vdash\psi\times\neg\phi$ and $\psi\times\neg\phi\vdash\psi\times\phi$.
\paragraph{The empty universe.}
Theorem~\ref{DBL:theo:1} implies $\neg\phi\vdash\psi\times\neg\phi$.\\
By applying b5.weak.A and CUT, it comes $\neg\phi\vdash\psi\times\phi$.\\[5pt]
The remaining proof is obvious.
\paragraph{Left equivalences.}
\emph{Proof of the main theorem.}\\[5pt]
Sequent $\psi\rightarrow\eta\vdash\phi\rightarrow(\psi\rightarrow\eta)$ is deduced classically.\\
Now, axiom b1 implies $\phi\rightarrow(\psi\rightarrow\eta)\vdash\neg\phi,(\psi\rightarrow\eta|\phi)$.\\
And by CUT, $\psi\rightarrow\eta\vdash\neg\phi,(\psi\rightarrow\eta|\phi)$\,.\\
Now, b2, modus ponens and CUT implies $(\psi\rightarrow\eta|\phi)\vdash(\psi|\phi)\rightarrow(\eta|\phi)$\,.\\
Applying CUT agains, it comes $\psi\rightarrow\eta\vdash\neg\phi,(\psi|\phi)\rightarrow(\eta|\phi)$\,.\\
Since $\psi$ and $\eta$ are exchangeable, the theorem is deduced.
\\

\emph{Proof of the corollary.}\\[5pt]
From~\ref{DBL:theo:3}, it comes $\neg\phi\vdash(\psi|\phi)\leftrightarrow\psi$ and $\neg\phi\vdash(\eta|\phi)\leftrightarrow\eta$.\\
Then $\neg\phi,\psi\leftrightarrow\eta\vdash(\psi|\phi)\leftrightarrow(\eta|\phi)$ by classical deductions.\\ 
The corollary is deduced by CUT with $\psi\rightarrow\eta\vdash\neg\phi,(\psi|\phi)\rightarrow(\eta|\phi)$ and STRUCT.
\paragraph{Sub-universes are classical.}
The first theorem is a consequence of axiom b4.
\vspace{5pt}\\
From axiom b2\,, it is deduced $\vdash(\psi\rightarrow\neg\eta|\phi)\rightarrow\bigl((\psi|\phi)\rightarrow(\neg\eta|\phi)\bigr)$\,.\\
It is deduced $\vdash\bigl((\psi|\phi)\wedge\neg(\neg\eta|\phi)\bigr)\rightarrow\neg(\psi\rightarrow\neg\eta|\phi)$\,.\\
Applying b4, it comes $\vdash\bigl((\psi|\phi)\wedge(\eta|\phi)\bigr)\rightarrow(\psi\wedge\eta|\phi)$.
\\[5pt]
Conversely, $\vdash\phi\rightarrow\bigl((\psi\wedge\eta)\rightarrow\psi\bigr)$ and b1 imply $\vdash\neg\phi,\bigl((\psi\wedge\eta)\rightarrow\psi\big|\phi\bigr)$.\\
By b2 it is deduced $\vdash\neg\phi,(\psi\wedge\eta|\phi)\rightarrow(\psi|\phi)$.\\
It is similarly proved $\vdash\neg\phi,(\psi\wedge\eta|\phi)\rightarrow(\eta|\phi)$.\\
As a consequence, $\vdash\neg\phi,(\psi\wedge\eta|\phi)\rightarrow\bigl((\psi|\phi)\wedge(\eta|\phi)\bigr)$.\\[3pt]
Moreover,~\ref{DBL:theo:3} implies $\neg\phi\vdash(\Xi|\phi)\leftrightarrow\Xi$ for $\Xi=\psi$, $\eta$ or $\psi\wedge\eta$.\\
It comes $\neg\phi\vdash(\psi\wedge\eta|\phi)\rightarrow\bigl((\psi|\phi)\wedge(\eta|\phi)\bigr)$ by a classical deduction.\\
By CUT on the derived sequents and STRUCT, it is deduced $\vdash(\psi\wedge\eta|\phi)\rightarrow\bigl((\psi|\phi)\wedge(\eta|\phi)\bigr)$.\\
The second theorem is then proved.
~\vspace{5pt}\\
Third theorem is a consequence of the first and second theorems.
~\vspace{5pt}\\
Last theorem is a consequence of the first and third theorems.
\paragraph{Evaluating $(\top|\cdot)$ and $(\bot|\cdot)$\,.}
From b1, it comes $\phi\rightarrow\psi\vdash\neg\phi,(\psi|\phi)$\,.\\
From $\vdash\psi\rightarrow(\phi\rightarrow\psi)$\,, modus ponens and CUT, it comes $\psi\vdash\phi\rightarrow\psi$\,.\\
Then $\psi\vdash\neg\phi,(\psi|\phi)$ by CUT\,.\\
Now $\neg\phi\vdash(\psi|\phi)\leftrightarrow\psi$ by~\ref{DBL:theo:3},
resulting in $\neg\phi\vdash\psi\rightarrow(\psi|\phi)$ and $\neg\phi,\psi\vdash(\psi|\phi)$.\\
Then $\psi\vdash(\psi|\phi)$\,, as a consequence of CUT and STRUCT.
\paragraph{Inference property.}
From b3 it comes $\vdash(\neg\psi|\phi)\rightarrow(\phi\rightarrow\neg\psi)$\,.\\
Then $\vdash\neg(\phi\rightarrow\neg\psi)\rightarrow\neg(\neg\psi|\phi)$ and 
$\vdash(\phi\wedge\psi)\rightarrow(\psi|\phi)$ by using b4.\\
At last $\vdash(\phi\wedge\psi)\rightarrow\bigl((\psi|\phi)\wedge\phi\bigr)$\,.\\[5pt]
Conversely $\vdash(\psi|\phi)\rightarrow(\phi\rightarrow\psi)$ implies $\vdash\bigl((\psi|\phi)\wedge\phi\bigr)\rightarrow\bigl((\phi\rightarrow\psi)\wedge\phi\bigr)$\,.\\
Since $(\phi\rightarrow\psi)\wedge\phi\equiv\phi\wedge\psi$\,, the converse is proved.
\paragraph{Introspection.}
Obvious from $\vdash \phi\rightarrow\phi$, b1 and CUT\,.
\paragraph{Inter-independence.}
It is proved:
$$
\bigl((\psi|\phi)\big|\phi\bigr)\wedge(\phi|\phi)
\equiv\bigl((\psi|\phi)\wedge\phi\big|\phi\bigr)\equiv(\phi\wedge\psi|\phi)\equiv(\psi|\phi)\wedge(\phi|\phi)\;.
$$
As a consequence $\vdash(\phi|\phi)\rightarrow\Bigl(\bigl((\psi|\phi)\big|\phi\bigr)\leftrightarrow(\psi|\phi)\Bigr)$\,, and then $(\phi|\phi)\vdash(\psi|\phi)\times\phi$\,.\\
Now $\vdash\neg\phi,(\phi|\phi)$ by~\ref{DBL:theo:8} and $\neg\phi\vdash(\psi|\phi)\times\phi$ as an instance of~\ref{DBL:theo:3}.\\
At last $\vdash(\psi|\phi)\times\phi$\,, by applying CUT and STRUCT.
\paragraph{Independence invariance.}
Since $(\psi|\phi)\leftrightarrow\psi\equiv\neg(\psi|\phi)\leftrightarrow\neg\psi\equiv(\neg\psi|\phi)\leftrightarrow\neg\psi$, it comes $\psi\times\phi\vdash\neg\psi\times\phi$\,.
\\[9pt]
Obviously, $(\psi|\phi)\leftrightarrow\psi,(\eta|\phi)\leftrightarrow\eta\vdash\bigl((\psi|\phi)\wedge(\eta|\phi)\bigr)\leftrightarrow(\psi\wedge\eta)$\,.\\
Applying~\ref{DBL:theo:5}, it comes $\psi\times\phi,\eta\times\phi\vdash(\psi\wedge\eta)\times\phi$\,.
\\[9pt]
From $\vdash(\alpha\leftrightarrow\beta)\rightarrow\Bigl((\gamma\leftrightarrow\delta)\rightarrow\bigl((\alpha\leftrightarrow\gamma)\rightarrow(\beta\leftrightarrow\delta)\bigr)\Bigr)$ is deduced
\gonextline
$\alpha\leftrightarrow\beta,\gamma\leftrightarrow\delta,\alpha\leftrightarrow\gamma\vdash\beta\leftrightarrow\delta\,.$\\
Now, \ref{DBL:theo:4} implies $\psi\leftrightarrow\eta\vdash(\psi|\phi)\leftrightarrow(\eta|\phi)$.\\
By replacing $\alpha,\beta,\gamma,\delta$ by $\psi,\eta,(\psi|\phi),(\eta|\phi)$ respectively, and applying CUT, it comes
\gonextline
$\psi\leftrightarrow\eta,\psi\times\phi\vdash\eta\times\phi$\,.
\paragraph{Narcissistic independence.}
From \ref{DBL:theo:8} comes $\vdash\neg\phi,(\phi|\phi)$\,.\\
From definition comes $\vdash(\phi\times\phi)\rightarrow\big((\phi|\phi)\rightarrow\phi\big)$ and then 
$\phi\times\phi,(\phi|\phi)\vdash\phi$.\\
Applying CUT, it is thus deduced $\phi\times\phi\vdash\neg\phi,\phi$\,.
\paragraph{Independence and proof.}
From $\vdash(\phi\vee\psi)\rightarrow(\neg\phi\rightarrow\psi)$ is deduced $\phi\vee\psi\vdash\neg\phi\rightarrow\psi$\,.\\
From b1 comes $\neg\phi\rightarrow\psi\vdash\phi,(\psi|\neg\phi)$\,.\\
Applying CUT then yields $\phi\vee\psi\vdash\phi,(\psi|\neg\phi)$\,.\\
From b5.weak.A, it comes $\psi\times\phi\vdash(\psi|\neg\phi)\leftrightarrow\psi$\,, and then  $\psi\times\phi,(\psi|\neg\phi)\vdash\psi$\,.\\
Applying CUT, it is deduced $\psi\times\phi,\phi\vee\psi\vdash\phi,\psi$\,.
\paragraph{Independence and regularity.}
\emph{Proof of the main theorem.}\\[5pt]
It is proved classically that $(\phi\wedge\eta)\rightarrow(\psi\wedge\eta)\equiv\neg\eta\vee(\phi\rightarrow\psi)$\,.\\
Then $(\phi\wedge\eta)\rightarrow(\psi\wedge\eta)\vdash\neg\eta\vee(\phi\rightarrow\psi)$\,.\\
Now, it is deduced $\phi\times\eta,\psi\times\eta\vdash(\phi\rightarrow\psi)\times\neg\eta$ from~\ref{DBL:theo:10} and b5.weak.A.\\
The proof is achieved by means of~\ref{DBL:theo:12}
and CUT.
\\[9pt]
Corollary is proved by applying CUT.
\\[9pt]
\emph{Proof of Corollary 2.}\\[5pt]
Assume $\vdash X\times\phi$ and $\neg\phi\vdash$\,.\\
Since $\vdash(\psi|\phi)\times\phi$ and $\psi\wedge\phi\equiv(\psi|\phi)\wedge\phi$, it is deduced from $X\wedge\phi\equiv\psi\wedge\phi$ that $X\equiv(\psi|\phi)$.
\paragraph{Right equivalences.}
First notice that all previous properties, \ref{DBL:theo:1} to \ref{DBL:theo:13}, are obtained without b5.weak.B.
\\[5pt]
From $(\phi|\psi)\wedge\psi\equiv\phi\wedge\psi$, $(\phi|\eta)\wedge\eta\equiv\phi\wedge\eta$ and $\vdash(\psi\leftrightarrow\eta)\rightarrow\bigl((\phi\wedge\psi)\leftrightarrow(\phi\wedge\eta)\bigr)$, it is  \gonextline deduced
$\vdash(\psi\leftrightarrow\eta)\rightarrow\Bigl(\bigl((\phi|\psi)\wedge\psi\bigr)\leftrightarrow\bigl((\phi|\eta)\wedge\eta\bigr)\Bigr)\,.$\\
Then $\vdash(\psi\leftrightarrow\eta)\rightarrow\Bigl(\bigl((\phi|\psi)\wedge\psi\bigr)\leftrightarrow\bigl((\phi|\eta)\wedge\psi\bigr)\Bigr)$ and finally
\gonextline
$
\psi\leftrightarrow\eta\vdash\bigl((\phi|\psi)\wedge\psi\bigr)\leftrightarrow\bigl((\phi|\eta)\wedge\psi\bigr)\ [a]\,.
$\\
Now $\vdash(\phi|\psi)\times\psi\ [b]$ and $\vdash(\phi|\eta)\times\eta$\,.
\\
Since $\psi\leftrightarrow\eta,\eta\times(\phi|\eta)\vdash\psi\times(\phi|\eta)$ by \ref{DBL:theo:5}, it comes $\psi\leftrightarrow\eta\vdash(\phi|\eta)\times\psi\ [c]$ by b5, CUT.
\\
Now, \ref{DBL:theo:13} implies $(\phi|\psi)\times\psi,
(\phi|\eta)\times\psi,
\bigl((\phi|\psi)\wedge\psi\bigr)\leftrightarrow\bigl((\phi|\eta)\wedge\psi\bigr)
\vdash\neg\psi,(\phi|\psi)\leftrightarrow(\phi|\eta)$.
\\
Combining it with $[a]$, $[b]$ and $[c]$ by CUT, it is obtained
$\psi\leftrightarrow\eta\vdash\neg\psi,(\phi|\psi)\leftrightarrow(\phi|\eta)\ [d]$\,.
\\[5pt]
Now, \ref{DBL:theo:3} implies $\neg\eta\vdash(\phi|\eta)\leftrightarrow\phi$\,.\\
It is easily proved $\psi\leftrightarrow\eta,\neg\psi\vdash\neg\eta$, and it is deduced $\neg\psi,\psi\leftrightarrow\eta\vdash(\phi|\eta)\leftrightarrow\phi$ by CUT.\\
Again, \ref{DBL:theo:3} yields $\neg\psi\vdash(\phi|\psi)\leftrightarrow\phi$, which combined with the previous sequent implies
\gonextline
$\neg\psi,\psi\leftrightarrow\eta\vdash(\phi|\eta)\leftrightarrow(\phi|\psi)\ [e]$.
\\
At last, $\psi\leftrightarrow\eta\vdash(\phi|\eta)\leftrightarrow(\phi|\psi)$ is obtained from $[d]$ and $[e]$ by CUT and STRUCT.
\paragraph{Reduction rule.}
$\vdash(\psi|\phi)\times\phi$ from \ref{DBL:theo:9} and $(\psi|\phi)\times\phi\vdash\phi\times(\psi|\phi)$ from b5 yield $\vdash\phi\times(\psi|\phi)$ by CUT.\\
Then $\bigl(\phi\big|(\psi|\phi)\bigr)\equiv\phi$ is just obtained as the definition of $\vdash\phi\times(\psi|\phi)$\,.
\paragraph{Markov Property.}
$\vdash(\phi_t|\phi_{t-1})\times\phi_{t-1}$ from \ref{DBL:theo:9}, and then, it is easily derived 
\gonextline
$(\phi_t|\phi_{t-1})\times\phi_{1},\dots,(\phi_t|\phi_{t-1})\times\phi_{t-2}\vdash(\phi_t|\phi_{t-1})\times\bigwedge_{\tau=1}^{t-1}\phi_{\tau}$\,.\\
Now, $(\phi_t|\phi_{t-1})\wedge \left(\bigwedge_{\tau=1}^{t-1}\phi_\tau\right)\equiv\bigwedge_{\tau=1}^{t}\phi_\tau
\equiv\left(\phi_t\left|\bigwedge_{\tau=1}^{t-1}\phi_\tau\right.\right)\wedge \left(\bigwedge_{\tau=1}^{t-1}\phi_\tau\right)$\,.\\
Since $\vdash\left(\phi_t\left|\bigwedge_{\tau=1}^{t-1}\phi_\tau\right.\right)\times \left(\bigwedge_{\tau=1}^{t-1}\phi_\tau\right)$\,, the proof is achieved by applying~\ref{DBL:theo:13}.
\paragraph{Link between $\bigl((\eta|\psi)\big|\phi\bigr)$ and $(\eta|\phi\wedge\psi)$.}
\emph{Proof of the logical counterpart to Lewis' triviality.}
\\[5pt]
It is equivalent to prove both $\vdash\neg(\phi\wedge\psi),\phi\rightarrow\psi,\phi\times\psi$ and $\vdash\neg(\phi\wedge\psi),\psi\rightarrow\phi,\phi\times\psi$.\\
Since $\times$ is symmetric, it is sufficient to prove $\vdash\neg(\phi\wedge\psi),\phi\rightarrow\psi,\phi\times\psi$.
\\[5pt]
Therorem~\ref{DBL:theo:8} implies $\vdash\neg(\phi\wedge\psi),(\phi\wedge\psi|\phi\wedge\psi)\ [a]$ and $\vdash\phi\rightarrow\psi,(\neg\psi\wedge\phi|\neg\psi\wedge\phi)$\,.\\
It is deduced $\vdash\neg(\phi\wedge\psi),(\psi|\psi\wedge\phi)\leftrightarrow\bigr((\psi|\psi\wedge\phi)\wedge(\psi\wedge\phi|\psi\wedge\phi)\bigl)\ [b]$ and
\gonextline
$\vdash\phi\rightarrow\psi,(\psi|\neg\psi\wedge\phi)\leftrightarrow\bigr((\psi|\neg\psi\wedge\phi)\wedge(\neg\psi\wedge\phi|\neg\psi\wedge\phi)\bigl)$.
\\
From the deduction $(\psi|\neg\psi\wedge\phi)\wedge(\neg\psi\wedge\phi|\neg\psi\wedge\phi)\equiv(\bot|\neg\psi\wedge\phi)\equiv\bot$, it is derived
\gonextline
$\vdash\phi\rightarrow\psi,(\psi|\neg\psi\wedge\phi)\leftrightarrow\bot\ [c]$.
\\
From the deduction $(\psi|\psi\wedge\phi)\wedge(\psi\wedge\phi|\psi\wedge\phi)\equiv(\psi\wedge\phi|\psi\wedge\phi)$, $[a]$ and $[b]$, it comes
\gonextline
$\vdash\neg(\phi\wedge\psi),(\psi|\psi\wedge\phi)\leftrightarrow\top\ [d]$.
\\
Now $(\psi|\phi)\equiv\bigl((\psi|\phi)\wedge\psi\bigr)\vee\bigl((\psi|\phi)\wedge\neg\psi\bigr)\equiv
\Bigl(\bigl((\psi|\phi)\big|\psi\bigr)\wedge\psi\Bigr)\vee\Bigl(\bigl((\psi|\phi)\big|\neg\psi\bigr)\wedge\neg\psi\Bigr)$\,, and
\gonextline
by applying axiom $(\ast)$, $(\psi|\phi)\equiv\bigl((\psi|\phi\wedge\psi)\wedge\psi\bigr)\vee\bigl((\psi|\phi\wedge\neg\psi)\wedge\neg\psi\bigr)$.
\\
Then $\vdash\phi\rightarrow\psi,\neg(\phi\wedge\psi),
(\psi|\phi)\leftrightarrow\bigl((\top\wedge\psi)\vee(\bot\wedge\neg\psi)\bigr)$ by means of $[c]$ and $[d]$.
\\
At last  $\vdash\phi\rightarrow\psi,\neg(\phi\wedge\psi),(\psi|\phi)\leftrightarrow\psi$\,.
\section{Proof: soundness of the semantic}
\label{Apx:Proof2Sem:sect}
The proof of proposition~\ref{DBL:prop:modelsound:1} is made recursively.
It is first proved that any axiom of DBL (resp. DBL$_\ast$) is true for any conditional model.
Then it is proved that the rules STRUCT and CUT are compliant with any conditional model.
\paragraph{Axiom modus ponens.}
Being assumed $H(\phi)=\Omega$ and $\sim H(\phi)\cup H(\psi)=\Omega$\,, it is implied $H(\psi)=\Omega$.
\\[5pt]
As a consequence, $\phi,\phi\rightarrow\psi\models_{\mathbf{M}}\psi$\,.
\paragraph{Axiom c1 to c3.}
Immediate since $(B,\cup,\cap,\sim,\emptyset,\Omega)$ is a Boolean algebra.
\paragraph{Axiom b1.}
Assume $H(\phi\rightarrow\psi)=\Omega$\,.
\\
Then $\sim H(\phi)\cup H(\psi)=\Omega$ and, consequently, $H(\phi)\subset H(\psi)$\,.
\\
Since $\mathbf{M}$ fulfilles $\beta1$\,, it comes $\sim H(\phi)=\Omega$ or $f\bigl(H(\psi),H(\phi)\bigr)=\Omega$\,.
\\
As a consequence, $H(\neg\phi)=\Omega$ or $H\bigl((\psi|\phi)\bigr)=\Omega$\,.
\\[5pt]
It is deduced $\phi\rightarrow\psi\models_{\mathbf{M}}\neg\phi,(\psi|\phi)$\,.
\paragraph{Axiom b2.}
Deduced from $\beta2$ and $\beta4$\,.
\paragraph{Axiom b3.}
Deduced from $\beta3$.
\paragraph{Axiom b4.}
Immediate from $\beta4$.
\paragraph{Axiom b5.}
Applying $\beta5$, it is deduced that $H\bigl((\psi|\phi)\bigr)=H(\psi)$ implies $H\bigl((\phi|\psi)\bigr)=H(\phi)$\,.
\\
Now, it is noticed that $H(\phi)=H(\psi)$
if and only if
$H(\phi\leftrightarrow\psi)=\Omega$\,.
\\
As a consequence, $H(\psi\times\phi)=\Omega$ implies $H(\phi\times\psi)=\Omega$\,.
\\[5pt]
Then, $\psi\times\phi\models_{\mathbf{M}}\phi\times\psi$\,.
\paragraph{Axiom b5.weak.A}
Assume $H\bigl((\psi|\phi)\leftrightarrow\psi\bigr)=\Omega$\,.
\\
Then $H\bigl((\psi|\phi)\bigr)=H(\psi)$\,, and $f\bigl(H(\psi), H(\phi)\bigr)=H(\psi)$\,.
\\
By applying $\beta5w$, it comes $f\bigl(H(\psi), \sim H(\phi)\bigr)=H(\psi)$\,.
\\
It follows $H\bigl((\psi|\neg\phi)\bigr)=H(\psi)$ and finally $H\bigl((\psi|\neg\phi)\leftrightarrow\psi\bigr)=\Omega$\,.
\\[5pt]
As a consequence,
$\psi\times\phi\models_{\mathbf{M}}\psi\times\neg\phi$\,.
\\[5pt]
$\psi\times\neg\phi\models_{\mathbf{M}}\psi\times\phi$ is an immediate corollary.
\paragraph{Axiom b5.weak.B}
Assume $H(\psi\leftrightarrow\eta)=\Omega$\\
Then, it is implied $H(\psi)=H(\eta)$\,.
\\
As a consequence, $f\bigl(H(\phi),H(\psi)\bigr)=f\bigl(H(\phi),H(\eta)\bigr)$\,.
\\
Then, $H\bigl((\phi|\psi)\bigr)=H\bigl((\phi|\eta)\bigr)$ and $H\bigl((\phi|\psi)\leftrightarrow(\phi|\eta)\bigr)=\Omega$\,.
\\[5pt]
Consequently, $\psi\leftrightarrow\eta\models_{\mathbf{M}}(\phi|\psi)\leftrightarrow(\phi|\eta)$\,.
\paragraph{Rule CUT.}
Immediate.
\paragraph{Rule STRUCT.}It is recalled that $\top=\theta_1\rightarrow\theta_1$ and $\bot=\neg\top$\,, where $\theta_1\in\Theta$\,.\\
As a consequence, $H(\top)=\sim H(\theta_1)\cup H(\theta_1)=\Omega$ and $H(\bot)=\sim H(\top)=\emptyset\ne\Omega$\,.\\
Then, $\{\Gamma\}\subset\{\Lambda\}\cup\{\top\}$\,, $\{\Delta\}\subset\{\Sigma\}\cup\{\bot\}$ and $\Gamma\models_{\mathbf{M}}\Delta$ imply $\Lambda\models_{\mathbf{M}}\Sigma$\,.
\section{Proof: properties of $(\mathbf{B}_{n},\cup,\cap,\sim,\emptyset,\Omega_{n}, f_{n}, \mu_n)_{n\in\Nset}$}
\label{Appendix:MainProof}
To be proved:
\emph{\begin{description}
\item[$\rien\quad\alpha1$] $\mu_n:\mathbf{B}_n\rightarrow \mathbf{B}_{n+1}$ is a one-to-one Boolean morphism\,,
\item[$\rien\quad\alpha2$] $f_{n+1}\bigl(\mu_n(B),\mu_n(A)\bigr)=\mu_n\bigl(f_n(B,A)\bigr)$\,,
\item[$\rien\quad\beta1$.] $A\subset B$ and $A\ne\emptyset$ imply $f_n(B,A)=\Omega_n$\,,
\item[$\rien\quad\beta2$.] $f_n(B\cup C,A)= f_n(B,A)\cup f_n(C,A)$\,,
\item[$\rien\quad\beta3$.] $A\cap f_n(B,A) = A\cap B$\,,
\item[$\rien\quad\beta4$.] $f_n(\sim B,A)=\sim f_n(B,A)$\,,
\item[$\rien\quad\beta5w$.] $f_n(B,A)=B$ implies $f_n(B,\sim A)=B$\,,
\end{description}
being assumed $A,B,C\in \mathbf{B}_n$\,, and $f_n(\cdot,\cdot)$ defined for the considered cases.
}
\\[5pt]
The proof is recursive and needs to consider the two cases of the definition of $(\mu_n,f_n)$\,.
\\[3pt]
The properties $\beta\ast$ are obvious for $n=0$, since $f_0$ is only defined by $f_0(A,\emptyset)=f_0(A,\Omega_0)=A$\,.
From now on, it is assumed that $\beta\ast$ hold true for $k\le n$, and that $\alpha1$ and $\alpha2$ hold true for $k\le n-1$\,.
The subsequent paragraphs establish the proof of $\beta\ast$ for $n+1$ and the proof of $\alpha1$ and $\alpha2$ for $n$.
\paragraph{Preliminary remarks.}
It is noticed that $\beta2$ and $\beta4$ imply:
$$\beta6:\ f_n(B\cap C,A)= f_n(B,A)\cap f_n(C,A)\,.$$
By construction, it is noticed that $f_n(\emptyset,A)=\emptyset$, when $f_n(\emptyset,A)$ exists.
\subsection{Lemma.}
\label{lemma:include}
$\bigcup_{i\in I_n}\Pi_n(i)=b_n$
and
$\bigcup_{i\in I_n}\Gamma_n(i)=\sim b_n$\,;
in particular, $\Pi_n(i)\cap\Gamma_n(j)=\emptyset$ for any $i,j\in I_n$\,.\\[5pt]
Moreover $\Pi_n(i)\cap\Pi_n(j)=\Gamma_n(i)\cap\Gamma_n(j)=\emptyset$ for any $i,j\in I_n$ such that $i\ne j$\,.
\begin{description}
\item[Proof.]The proof is obvious for case 1.\\
Now, let consider case 0.\\
By definition $
\bigcup_{i\in I_n}\Pi_n(i)=\left(\bigcup_{\omega\in\mu_\nu(b_\nu)}\omega_{[n}\right)
\cap
\left(\bigcup_{\omega'\in\sim\mu_\nu(b_\nu)}f_n(\omega'_{[n},\sim b_n)\right)
$\,.\\
By recursion hypothesis $\beta1$, it comes $f_n(\sim b_n,\sim b_n)=\Omega_n$\,.
\\
Since $b_n=b_{\nu[n}$ and by $\beta2$ and $\alpha_1$\,, it comes \gonextline$\bigcup_{i\in I_n}\Pi_n(i)=b_n\cap f_n(\sim b_n,\sim b_n)=b_n\cap\Omega_n= b_n\;.$
\\[5pt]
For any $\omega_1,\omega_2\in \sim\mu_\nu(b_\nu)$ such that $\omega_1\ne\omega_2$\,, it comes by $\beta6$ (\emph{i.e.} by $\beta2$ and $\beta4$)\,:
$$
f_n(\omega_{1[n},\sim b_n)\cap f_n(\omega_{2[n},\sim b_n)=f_n(\omega_{1[n}\cap\omega_{2[n},\sim b_n)=f_n(\emptyset,\sim b_n)=\emptyset\;.
$$
Finally $\Pi_n(i)\cap\Pi_n(j)=\emptyset$ for any $i,j\in I_n$ such that $i\ne j$\,.\\[5pt]
The results are similarly proved for $\Gamma_n$\,.
\item[$\Box\Box\Box$]\rien
\end{description}
\emph{Corollary 1.}
$$
\mu_n(b_n)=T(\sim\mu_n(b_n))=\bigcup_{i\in I_n}\Pi_n(i)\times\Gamma_n(i)
\quad\mbox{and}\quad
\sim\mu_n(b_n)=T(\mu_n(b_n))=\bigcup_{i\in I_n}\Gamma_n(i)\times\Pi_n(i)\;.
$$
\emph{Corollary 2.}
$$f_{n+1}\bigl(C,\mu_n(b_n)\bigr)=\bigl(C\cap\mu_n(b_n)\bigr)\cup\bigl(T(C)\cap\sim\mu_n(b_n)\bigr)$$
and
$$f_{n+1}\bigl(C,\sim\mu_n(b_n)\bigr)=\bigl(T(C)\cap\mu_n(b_n)\bigr)\cup\bigl(C\cap\sim\mu_n(b_n)\bigr)\;.$$
Both corollaries are obvious from the definition.
\subsection{Useful set properties}
\label{Proof:Setprop}
The following properties (whose proofs are immediate) will be useful:
\begin{description}
\item[$\ell1.$] $(A\cup B)\times C=(A\times C)\cup (B\times C)$
and $A\times (B\cup C)=(A\times B)\cup (A\times C)$\,,
\item[$\ell2.$] $(A\cap B)\times C=(A\times C)\cap (B\times C)$
and $A\times (B\cap C)=(A\times B)\cap (A\times C)$\,,
\item[$\ell3.$] $C\cap D=\emptyset$ implies $(C\times A)\cap(D\times B)=(A\times C)\cap(B\times D)=\emptyset$\,,
\item[$\ell4.$] $(A\cup B)\cap (C\cup D)=\emptyset$ implies $(A\cap B)\cup(C\cap D)=(A\cup C)\cap(B\cup D)$\,,
\item[$\ell5.$] $(A\cup B)\cap (C\cup D)=\emptyset$ and $A\cup C=B\cup D$ imply $A=B$ and $C=D$\,,
\item[$\ell6.$] $C\cap D=\emptyset$ implies $(C\cup D)\setminus\bigl((A\cap C)\cup(B\cap D)\bigr)=(C\setminus A)\cup(D\setminus B)$\,,
\end{description}
for any sets $A,B,C,D$\,.
\subsection{Proof of $\alpha1$}
\label{Proof:Cap}
\paragraph{\underline{Proof of $\mu_n(\Omega_n)=\Omega_{n+1}$ and $\mu_n(\emptyset)=\emptyset$\,.}}
Immediate from the definitions.
\paragraph{\underline{Proof of $\mu_n(A\cap B)=\mu_n(A)\cap\mu_n(B)$\,.}}
By applying $\ell2$, it is deduced:
$$\begin{array}{@{}l@{}}
\mu_n(A\cap B)=\bigcup_{i\in I_n} \biggl(\Bigl(\bigl(A\cap\Pi_n(i)\bigr)\times\Gamma_n(i)\Bigr)
\cap
\Bigl(\bigl(B\cap\Pi_n(i)\bigr)\times\Gamma_n(i)\Bigr)\biggr)
\vspace{5pt}\\\rien\hspace{50pt}
\cup\ 
\bigcup_{i\in I_n}
\biggl(\Bigl(\bigl(A\cap\Gamma_n(i)\bigr)\times\Pi_n(i)\Bigr)
\cap
\Bigl(\bigl(B\cap\Gamma_n(i)\bigr)\times\Pi_n(i)\Bigr)\biggr)
\;.
\end{array}$$
By lemma~\ref{lemma:include}, and applying $\ell3$ and $\ell4$, it is deduced:
$$\begin{array}{@{}l@{}}
\mu_n(A\cap B)=\bigcup_{i\in I_n} \biggl(\Bigl(\bigl(A\cap\Pi_n(i)\bigr)\times\Gamma_n(i)\Bigr)
\cup
\Bigl(\bigl(A\cap\Gamma_n(i)\bigr)\times\Pi_n(i)\Bigr)\biggr)
\vspace{5pt}\\\rien\hspace{50pt}
\cap\ 
\bigcup_{i\in I_n}
\biggl(\Bigl(\bigl(B\cap\Pi_n(i)\bigr)\times\Gamma_n(i)\Bigr)
\cup
\Bigl(\bigl(B\cap\Gamma_n(i)\bigr)\times\Pi_n(i)\Bigr)\biggr)
=\mu_n(A)\cap\mu_n(B)\;.
\end{array}$$
\paragraph{\underline{Proof of $\mu_n(A\cup B)=\mu_n(A)\cup\mu_n(B)$\,.}}
Obviously deduced from $\ell1$.
\paragraph{\underline{$\mu_n$ is one-to-one.}}
Assume $\mu_n(A)=\mu_n(B)$\,;
then:
$$\begin{array}{@{}l@{}}
\bigcup_{i\in I_n} \biggl(\Bigl(\bigl(A\cap\Pi_n(i)\bigr)\times\Gamma_n(i)\Bigr)
\cup
\Bigl(\bigl(A\cap\Gamma_n(i)\bigr)\times\Pi_n(i)\Bigr)\biggr)
\vspace{5pt}\\\rien\hspace{50pt}
=
\bigcup_{i\in I_n}
\biggl(\Bigl(\bigl(B\cap\Pi_n(i)\bigr)\times\Gamma_n(i)\Bigr)
\cup
\Bigl(\bigl(B\cap\Gamma_n(i)\bigr)\times\Pi_n(i)\Bigr)\biggr)
\;.
\end{array}$$
By lemma~\ref{lemma:include}, and applying $\ell2$, $\ell3$ and $\ell5$, it is deduced for any $i\in I_n$\,:
$$
\bigl(A\cap\Pi_n(i)\bigr)\times\Gamma_n(i)
=
\bigl(B\cap\Pi_n(i)\bigr)\times\Gamma_n(i)
\ 
\mbox{ and }
\ 
\bigl(A\cap\Gamma_n(i)\bigr)\times\Pi_n(i)
=
\bigl(B\cap\Gamma_n(i)\bigr)\times\Pi_n(i)
\;.$$
Finally $A\cap\Pi_n(i)=B\cap\Pi_n(i)$ and $A\cap\Gamma_n(i)=B\cap\Gamma_n(i)$ for any $i\in I_n$\,, and:
$$
A\cap\bigcup_{i\in I_n}\bigl(\Pi_n(i)\cup\Gamma_n(i)\bigr)=B\cap\bigcup_{i\in I_n}\bigl(\Pi_n(i)\cup\Gamma_n(i)\bigr)\;.
$$
$A=B$ is deduced by applying the lemma.
\paragraph{Conclusion.} The previous results imply that $\mu_n$ is a one-to-one Boolean morphism.
\subsection{Proof of $\alpha2$}\label{Proof:idempot}
By definition, the result holds true for any $A\in M_n\setminus\{\emptyset,\Omega_n,b_n,\sim b_n\}$\,.
It is also true for $A=\emptyset$ or $A=\Omega_n$\,, since $f_{n+1}\bigl(\mu_n(B),\mu_n(\emptyset)\bigl)=f_{n+1}\bigl(\mu_n(B),\emptyset\bigl)=\mu_n(B)=\mu_n\bigl(f_n(B,\emptyset)\bigr)$
and similarly
$f_{n+1}\bigl(\mu_n(B),\mu_n(\Omega_n)\bigl)=f_{n+1}\bigl(\mu_n(B),\Omega_{n+1}\bigl)=\mu_n(B)=\mu_n\bigl(f_n(B,\Omega_n)\bigr)$\,.
\\[5pt]
The true difficulties come from the cases $A=b_n$ or $A=\sim b_n$\,.\\
Subsequently, it is assumed $A=b_n$\,; the case $A=\sim b_n$ is quite similar.\\[5pt]
It comes:
$$\begin{array}{@{}l@{}}\displaystyle
f_{n+1}\bigl(\mu_n(B),\mu_n(b_n)\bigr)=(\mathrm{id}\cup T)
\biggl(\mu_n(B)\cap\Bigl(\bigcup_{i\in I_n}\bigl(\Pi_n(i)\times\Gamma_n(i)\bigr)\Bigr)\biggr)
\hspace{100pt}\rien
\gonextline\displaystyle
=(\mathrm{id}\cup T)\Bigl(\bigcup_{i\in I_n}\bigl(B\cap\Pi_n(i)\bigr)\times\Gamma_n(i)\Bigr)\;.
\end{array}$$
The existence of $f_n(B,b_n)$ implies the case 0\,,
and there is $C\in M_{\nu+1}$ such that $B=C_{[n}$\,.
\\
By recursion hypotheses $\alpha1$, it comes:
$$B\cap\omega_{[n}=(C\cap\omega)_{[n}=\omega_{[n}\mbox{ if }\omega\in C\,,\quad
=\emptyset\mbox{ if }\omega\not\in C\,.$$
By recursion hypotheses $\beta6$ (\emph{i.e.} by $\beta2$ and $\beta4$), it comes:
$$f_n(\omega_{[n},b_n)\cap f_n(B,b_n)=f_n(\omega_{[n}\cap B,b_n)=f_n(\omega_{[n},b_n)\mbox{ if }\omega\in C\,, 
\quad=\emptyset\mbox{ if }\omega\not\in C\,.$$
As a consequence $\bigcup_{i\in I_n}\bigl(B\cap\Pi_n(i)\bigr)\times\Gamma_n(i)=\bigcup_{i\in I_n}\Pi_n(i)\times\bigl(f_n(B,b_n)\cap\Gamma_n(i)\bigr)$\,.
\\[3pt]
By $\beta3$,
$B\cap\omega_{[n}=B\cap b_n\cap\omega_{[n}=f_n(B,b_n)\cap b_n\cap\omega_{[n}=f_n(B,b_n)\cap\omega_{[n}$ for any $\omega\in\mu_\nu(b_\nu)$\,.\\
As a consequence $\bigcup_{i\in I_n}\bigl(B\cap\Pi_n(i)\bigr)\times\Gamma_n(i)=\bigcup_{i\in I_n}\bigl(f_n(B,b_n)\cap\Pi_n(i)\bigr)\times\Gamma_n(i)$\,.
\\
By applying the both results, it comes:
$$
f_{n+1}\bigl(\mu_n(B),\mu_n(b_n)\bigr)=
\Bigl(\bigcup_{i\in I_n}\bigl(f_n(B,b_n)\cap\Pi_n(i)\bigr)\times\Gamma_n(i)\Bigr)
\cup
\Bigl(\bigcup_{i\in I_n}\bigl(f_n(B,b_n)\cap\Gamma_n(i)\bigr)\times\Pi_n(i)\Bigr)\;.
$$
And by definition of $\mu_n$, it is finally deduced $f_{n+1}\bigl(\mu_n(B),\mu_n(b_n)\bigr)=\mu_n\bigl(f_{n}(B,b_n)\bigr)$\,.
\subsection{Proof of $\beta1$}
For $A\not\in \{\mu_n(b_n),\sim \mu_n(b_n),\emptyset,\Omega_{n+1}\}$, the propriety is inherited from $n$ by applying $\alpha2$\,.\\
The property is also obvious for $A\in \{\emptyset,\Omega_{n+1}\}$\,.
\\
The difficulty comes from $A=\mu_n(b_n)$ or $A=\sim \mu_n(b_n)$\,;
then notice that $A\ne\emptyset$ by construction.
\\[5pt]
It is now hypothesized $A=\mu_n(b_n)\subset B$\,;
the case $A=\sim \mu_n(b_n)$ is quite similar.\\
Then $T\bigl(\mu_n(b_n)\bigr)\subset T(B)$ and by lemma, corollary 1\&2:
$$
f_{n+1}\bigl(B,\mu_n(b_n)\bigr)=
\bigl(B\cap\mu_n(b_n)\bigr)\cup\bigl(T(B)\cap\sim\mu_n(b_n)\bigr)=\mu_n(b_n)\cup\sim\mu_n(b_n)
=\Omega_{n+1}\;.
$$
\subsection{Proof of $\beta2$}
For $A\not\in \{\mu_n(b_n),\sim \mu_n(b_n),\emptyset,\Omega_{n+1}\}$, the propriety is inherited from $n$ by applying $\alpha2$\,.\\
The property is also obvious for $A\in \{\emptyset,\Omega_{n+1}\}$\,.
\\
The property is immediate for $A\in\{\mu_n(b_n),\sim \mu_n(b_n)\}$\,, since $T(B_1\cup B_2)=T(B_1)\cup T(B_2)$.
\subsection{Proof of $\beta3$}
For $A\not\in \{\mu_n(b_n),\sim \mu_n(b_n),\emptyset,\Omega_{n+1}\}$, the propriety is inherited from $n$ by applying $\alpha2$\,.
The property is also obvious for $A\in \{\emptyset,\Omega_{n+1}\}$\,.
\\
The difficulty comes from $A\in \{\mu_n(b_n),\sim \mu_n(b_n)\}$.
\\[5pt]
It is now hypothesized $A=\mu_n(b_n)$\,;
the case $A=\sim \mu_n(b_n)$ is quite similar.\\
The result is immediate from corollary 2 of lemma.
\subsection{Proof of $\beta4$}
For $A\not\in \{\mu_n(b_n),\sim \mu_n(b_n),\emptyset,\Omega_{n+1}\}$, the propriety is inherited from $n$ by applying $\alpha2$\,.
The property is also obvious for $A\in \{\emptyset,\Omega_{n+1}\}$\,.
\\
The difficulty comes from $A\in \{\mu_n(b_n),\sim \mu_n(b_n)\}$\,.\\[5pt]
It is now hypothesized $A=\mu_n(b_n)$\,;
the case $A=\sim \mu_n(b_n)$ is quite similar.\\
By corollary 2 of lemma, $f_{n+1}\bigl(\sim B,\mu_{n}(b_n)\bigr)=\bigl(\mu_n(b_n)\setminus B\bigr)\cup\bigl(\sim\mu_n(b_n)\setminus T(B)\bigr)$\,.\\
By $\ell6$, $f_{n+1}\bigl(\sim B,\mu_{n}(b_n)\bigr)=
\sim\Bigl(\bigl(B\cap\mu_n(b_n)\bigr)\cup\bigl(T(B)\cap\sim\mu_n(b_n)\bigr)\Bigr)=\sim f_{n+1}\bigl(B,\mu_{n}(b_n)\bigr)$\,.
\subsection{Lemma 2.}
Let $C\in \mathbf{B}_{n+1}$\,. Then:
$$
f_{n+1}\Bigl(f_{n+1}\bigl(C,\mu_n(b_n)\bigr),\mu_n(b_n)\Bigr)=
f_{n+1}\Bigl(f_{n+1}\bigl(C,\mu_n(b_n)\bigr),\sim\mu_n(b_n)\Bigr)=
f_{n+1}\bigl(C,\mu_n(b_n)\bigr)
$$
and
$$
f_{n+1}\Bigl(f_{n+1}\bigl(C,\sim\mu_n(b_n)\bigr),\mu_n(b_n)\Bigr)=
f_{n+1}\Bigl(f_{n+1}\bigl(C,\sim\mu_n(b_n)\bigr),\sim\mu_n(b_n)\Bigr)=
f_{n+1}\bigl(C,\sim\mu_n(b_n)\bigr)
\;.
$$
\begin{description}
\item[Proof.]
The result is derived for $f_{n+1}\bigl(C,\mu_n(b_n)\bigr)$\,; it is quite similar for $f_{n+1}\bigl(C,\sim\mu_n(b_n)\bigr)$\,.
\\[3pt]
By corollary 2 of lemma, $f_{n+1}\bigl(C,\mu_n(b_n)\bigr)=\bigl(C\cap\mu_n(b_n)\bigr)\cup\bigl(T(C)\cap\sim\mu_n(b_n)\bigr)$\,.\\
Since $T\bigl(f_{n+1}\bigl(C,\mu_n(b_n)\bigr)\bigr)=f_{n+1}\bigl(C,\mu_n(b_n)\bigr)$ by definition, the proof is achieved by applying corollary 2.
\item[$\Box\Box\Box$]\rien
\end{description}
\emph{Corollary.}
As a direct consequence, $f_{n+1}\bigl(f_{n+1}(B,A),A\bigr)=f_{n+1}\bigl(f_{n+1}(B,A),\sim A\bigr)=f_{n+1}(B,A)$\,, whenever $f_{n+1}(B,A)$ exists.
\subsection{Proof of $\beta5w$}
Assume $f_{n+1}(B,A)$ and $f_{n+1}(B,\sim A)$ exist and $f_{n+1}(B,A)=B$\,.\\
By corollary of lemma 2, $f_{n+1}(B,\sim A)=f_{n+1}\bigl(f_{n+1}(B,A),\sim A\big)=f_{n+1}(B,A)=B$\,.
\section{Proof: completeness for the conditional operator}
\label{Appendix:ProofOfAlmostCompletude}
It is defined $h:\Theta\rightarrow\mathbf{B}[\Theta]$ by $h(\theta)=\nu_0(\xi_\theta)$, where 
$\xi_\theta=\bigl\{(\delta_\tau)_{\tau\in\Theta}\in\Omega_0\,\big/\,\delta_\theta=1\bigr\}$\,.
\\[5pt]
To be proved the equivalence of the assertions:
\begin{enumerate}
\item \label{dbl:comp:proof:1} $\vdash\phi$ in DBL$_\ast$\,,
\item \label{dbl:comp:proof:2} $\overline{h}(\phi)=\Omega$\,,
\item \label{dbl:comp:proof:3} $\models_{\mathbf{M}[\Theta]}\phi$\,.
\end{enumerate}
Notice that \ref{dbl:comp:proof:1} implies~\ref{dbl:comp:proof:3} and \ref{dbl:comp:proof:3} implies~\ref{dbl:comp:proof:2}\,.
As a consequence, it is sufficient to prove:
\begin{equation}\label{dbl:tobeproved:completude:1}
\rien\qquad\overline{h}(\phi)=\Omega\mbox{ implies }\vdash\phi\mbox{ in DBL}_\ast\;.
\end{equation}
Let $\mathcal{L}_\equiv$ be the factor set of $\mathcal{L}$ by the logical equivalence $\equiv$ of DBL$_\ast$.
\\
Define $\overline{h}_\equiv:\mathcal{L}_\equiv\rightarrow\mathbf{B}[\Theta]$ by:
$$
\overline{h}_\equiv(\phi_\equiv)=\bigl(\overline{h}(\phi)\bigr)_\equiv \quad\mbox{ for any }\phi\in\mathcal{L}\,.
$$
Then, proposition~(\ref{dbl:tobeproved:completude:1}) is a corollary of:
\begin{equation}
\label{the:Property}
\rien\qquad\mbox{$\overline{h}_\equiv$ is a Boolean isomorphism from $\mathcal{L}_\equiv$ to $\mathbf{B}[\Theta]$\,,}
\end{equation}
which will be derived from a recursive construction of $\mathcal{L}$ similar to the definition of $\mathbf{B}[\Theta]$\,.
\paragraph{Construction.}
Assume the sequence $(\mathbf{B}_{n},\cup,\cap,\sim,\emptyset,\Omega_{n}, f_{n}, \mu_n,b_n)_{n\in\Nset}$ being constructed.\\
The sequence $(L_n)_{n\in\Nset}$ is defined by:
\begin{itemize}
\item $L_0=\mathcal{L}_C$\,,
\item $L_{n+1}\subset\mathcal{L}$ is the set generated by $L_n$, the classical operators, the conditionals $(\cdot|\phi)$ and $(\cdot|\neg\phi)$ where $\phi$ is any proposition of $L_n$ such that $\overline{h}(\phi)=\nu_n(b_n)$\,.
\end{itemize}
A set $\Sigma_n\subset (L_n)_\equiv$ is called a generating partition of $(L_n)_\equiv$, if it verifies:
$$
\forall\phi\in (L_n)_\equiv\,,\;\exists S\subset\Sigma_n\,,\;\bigvee_{\sigma\in S}\sigma=\phi
\quad\mbox{and}\quad
\sigma\wedge\sigma'=\bot\mbox{ for any }\sigma,\sigma'\in\Sigma_n\mbox{ such that }\sigma\ne\sigma'\,.
$$
The following property is proved recursively in the next paragraphs:
\begin{equation}
\label{the:Equation}
\rien\qquad\mbox{There is a generating partition }\Sigma_n\mbox{ of }(L_n)_\equiv\mbox{ such that }\mathrm{card}(\Sigma_n)\le\mathrm{card}(\Omega_n)\;.
\end{equation}
Since $\overline{h}_\equiv$ is by construction an onto morphism from $(L_n)_\equiv$ to $\nu_n(\mathbf{B}_{n})$\,,
(\ref{the:Equation}) implies that $\overline{h}_\equiv$ is a Boolean isomorphism from $(L_n)_\equiv$ to $\nu_n(\mathbf{B}_{n})$\,.\\
The definition of $b_n$ implies the condition~(\ref{eq:hyp:proj:1}) of the direct limit for $\mathbf{B}_{n}$.
Then, for any $n\in\Nset$, also exists $m>n$ such that $(\psi|\phi)\in L_m$ for any $\phi,\psi\in L_n$\,.
\\[5pt]
As a consequence,
$\mathcal{L}_\equiv=\cup_{n\in\Nset}(L_n)_\equiv$ and~(\ref{the:Equation}) implies~(\ref{the:Property}).
\subsection{Recursive proof of (\ref{the:Equation})}
\paragraph{True for $n=0$.}
Obvious, since $\mathbf{B}_{0}$ is isomorph to the factor set of $\mathcal{L}_C$ by $\equiv_C$, and $\equiv_C$ is weaker than~$\equiv$.
\paragraph{True for $n$ implies true for $n+1$.}
The recursion hypothesis implies that $\overline{h}_\equiv$ is an isomorphism from $(L_n)_\equiv$ to $\nu_n(\mathbf{B}_{n})$\,.\\
Then, define $\beta_n\in (L_n)_\equiv$ such that $\overline{h}_\equiv(\beta_n)=b_n$ \emph{(notice that $\beta_n\ne\bot$ and $\neg\beta_n\ne\bot$)}.\\
It is known, from~\ref{DBL:theo:9} and~\ref{DBL:theo:10}, that: $$\bigl((\cdot|\beta_n)\big|\neg\beta_n\bigr)=\bigl((\cdot|\beta_n)\big|\beta_n\bigr)=(\cdot|\beta_n)
\mbox{ and }
\bigl((\cdot|\neg\beta_n)\big|\beta_n\bigr)=\bigl((\cdot|\neg\beta_n)\big|\neg\beta_n\bigr)=(\cdot|\neg\beta_n)
\;.$$
Then, by applying~\ref{DBL:theo:5}, it comes that $(L_{n+1})_\equiv$ is generated by: $$\Sigma_{n+1}=\bigl\{\sigma\wedge(\sigma'|\beta_n)\wedge(\sigma''|\neg\beta_n)\;/\;\sigma,\sigma',\sigma''\in\Sigma_n\bigr\}\setminus\{\bot\}\,.$$
Now, denote $B_n
=\bigl\{\sigma\in\Sigma_n\,\big/\,\sigma\wedge \beta_n=\sigma\bigr\}$
and
$\overline{B}_n
=\bigl\{\sigma\in\Sigma_n\,\big/\,\sigma\wedge\neg\beta_n=\sigma\bigr\}$\,.
\\
It comes from~b1 and~\ref{DBL:theo:5}, that $(\sigma'|\beta_n)=(\sigma''|\neg\beta_n)=\bot$ for $\sigma'\in\overline{B}_n$ and $\sigma''\in B_n$\,.\\
Moreover, from~\ref{DBL:theo:7}, $\sigma\wedge(\sigma'|\beta_n)\wedge(\sigma''|\neg\beta_n)=\bot$ for $\sigma\not\in \{\sigma',\sigma''\}$\,;
on the other hand,
$\sigma\wedge(\sigma|\beta_n)=\sigma$ for $\sigma\in B_n$\,, and
$\sigma\wedge(\sigma|\neg\beta_n)=\sigma$ for $\sigma\in \overline{B}_n$\,.
\\[5pt]
Then, the two construction cases of $(\mathbf{B}_{n},\cup,\cap,\sim, \emptyset,\Omega_{n}, f_{n}, \mu_n,b_n)_{n\in\Nset}$ are considered:
\subparagraph{Case 1.}
$\Sigma_{n+1}=
\bigcup_{\sigma\in B_n}
\bigcup_{\sigma'\in \overline{B}_n}\bigl\{\sigma\wedge(\sigma'|\neg\beta_n),\sigma'\wedge(\sigma|\beta_n)\bigr\}
$\,, owing to the above discussion.
\\
As a consequence, $
\mathrm{card}(\Sigma_{n+1})\le2\mathrm{card}(B_n)\mathrm{card}(\overline{B}_n)=2\mathrm{card}(b_n)\mathrm{card}(\sim b_n)=\mathrm{card}(\Omega_{n+1})\,.$
\subparagraph{Case 0.}
In this case, $\beta_n=\beta_\nu$\,.\\
Define $C_\nu=\{\sigma\in\Sigma_{\nu+1}/\sigma\wedge\beta_\nu=\sigma\}$
and $\overline{C}_\nu=\{\sigma\in\Sigma_{\nu+1}/\sigma\wedge\neg\beta_\nu=\sigma\}$\,.\\
Define also $D[\phi]=\{\sigma\in\Sigma_n/\sigma\wedge\phi=\sigma\}$ for any $\phi\in(L_{\nu+1})_\equiv$\,.\\[5pt]
From previously, it is know that $\Sigma_{n+1}$ contains elements of the form $\sigma\wedge(\sigma'|\neg\beta_n)$ or $\sigma'\wedge(\sigma|\beta_n)$ with $(\sigma,\sigma')\in B_n\times \overline{B}_n$\,;
but the construction at step $\nu+1$ implies additional constraints, to be specified.
\\[3pt]
Let consider the case $\sigma\wedge(\sigma'|\neg\beta_n)$ with $(\sigma,\sigma')\in B_n\times \overline{B}_n$\,; \emph{case $\sigma'\wedge(\sigma|\beta_n)$ is quite similar.}\\
Notice that there is $(\tau,\tau')\in C_\nu\times\overline{C}_\nu$
such that $\sigma\in D[\tau]\cap D[(\tau'|\neg\beta_\nu)]$\,,
and $(\theta,\theta')\in \overline{C}_\nu\times C_\nu$
such that $\sigma'\in D[\theta]\cap D[(\theta'|\beta_\nu)]$\,.\\
Now, $\tau\wedge(\tau'|\neg\beta_\nu)\wedge\bigl(\theta\wedge(\theta'|\beta_\nu)\big|\neg\beta_\nu\bigr)=
\bigl(\tau\wedge(\theta'|\beta_\nu)\bigr)\wedge
(\tau'\wedge\theta|\neg\beta_\nu)=\bot$
unless $\tau=\theta'$ and $\tau'=\theta$\,.\\
As a consequence, it is deduced:
$$\sigma\wedge(\sigma'|\neg\beta_n)\ne\bot\quad\mbox{implies}\quad\exists(\tau,\theta)\in C_\nu\times\overline{C}_\nu\,,\;(\sigma,\sigma')\in \bigl(D[\tau]\cap D[(\theta|\neg\beta_\nu)]\bigr)\times\bigl(D[\theta]\cap D[(\tau|\beta_\nu)]\bigr)\;.$$
Similarly, it is deduced:
$$\sigma'\wedge(\sigma|\beta_n)\ne\bot\quad\mbox{implies}\quad\exists(\theta,\tau)\in \overline{C}_\nu\times C_\nu\,,\;(\sigma',\sigma)\in \bigl(D[\theta]\cap D[(\tau|\beta_\nu)]\bigr)\times\bigl(D[\tau]\cap D[(\theta|\neg\beta_\nu)]\bigr)\;.$$
At last $\mathrm{card}(\Sigma_{n+1})\le\sum_{(\tau,\theta)\in C_\nu\times\overline{C}_\nu} 2\,\mathrm{card}\bigl(D[\tau]\cap D[(\theta|\neg\beta_\nu)]\bigr) \mathrm{card}\bigl(D[\theta]\cap D[(\tau|\beta_\nu)]\bigr)$
\gonextline
$=\sum_{i\in I_n}2\,\mathrm{card}\bigl(\Pi_n(i)\bigr) \mathrm{card}\bigl(\Gamma_n(i)\bigr)=\mathrm{card}(\Omega_{n+1})\;.$
\section{Probability extension}
\label{Appendix:Probabilition}
To be proved:\\[5pt]
Let $\pi$ be a probability defined over $C$\,, such that $\pi(\phi)>0$ for any $\phi\not\equiv_C\bot$.
Then, there is a (multiplicative) probability $\overline{\pi}$ defined over DBL$_\ast$ such that  $\forall\phi\in\mathcal{L}_C\,,\;\overline{\pi}(\phi)=\pi(\phi)$\,.
\\[5pt]
The construction of $\overline{\pi}$ is based on the recursive definition of $(\mathbf{B}_n,\cup,\cap,\sim,\emptyset,\Omega_n, f_n, b_n,\mu_n)_{n\in\Nset}$\,.
\subsection{Construction}
For any $n\in\Nset$, the probability $P_n$ is defined over $\mathbf{B}_n$ by:
$$
P_n(A)=\sum_{\omega\in A}P_n(\omega)\quad\mbox{for any }A\in \mathbf{B}_n\,,
$$
and:
\subparagraph{Initialization.}\rien\\
For $\omega=\bigl(\delta_\theta\bigr)_{\theta\in\Theta}\in\Omega_0$ and $\tau\in\Theta$, define:
\begin{equation}\label{dbl:apxF:deftauom:1}
\tau_\omega=\tau\mbox{ if }\delta_\tau=1\mbox{ and }\tau_\omega=\neg\tau\mbox{ if }\delta_\tau=0\,.
\end{equation}
Then set $P_0(\omega)=\pi\left(\bigwedge_{\tau\in\Theta}\tau_\omega\right)$ for any $\omega\in\Omega_0$\,.
\subparagraph{From n to n+1.}
For any $(\omega,\omega')\in\Pi_n(i)\times\Gamma_n(i)$\,, set:
$$
P_{n+1}(\omega,\omega')=\frac{P_n(\omega)P_n(\omega')}{P_n\bigl(\Gamma_n(i)\bigr)}
\quad\mbox{and}\quad
P_{n+1}(\omega',\omega)=\frac{P_n(\omega)P_n(\omega')}{P_n\bigl(\Pi_n(i)\bigr)}\;.
$$
In particular, $P_{n+1}\bigl(T(\omega)\bigr)=P_{n+1}(\omega)\frac{P_n\bigl(\Gamma_n(i)\bigr)}{P_n\bigl(\Pi_n(i)\bigr)}$ for $\displaystyle\omega\in \bigcup_{i\in I_n}\bigl(\Pi_n(i)\times\Gamma_n(i)\bigr)$\,.
%
%
\subsection{Lemmas.}
\subsubsection{Lemma 1} $P_{n+1}(A_{[n+1})=P_{n}(A)$ for any $A\in \mathbf{B}_n$\,.
\begin{description}
\item[Proof.]
For $A\in \mathbf{B}_n$\,,
$\mu_n(A)=\bigcup_{i\in I_n}\biggl(\Bigl(\bigl(A\cap\Pi_n(i)\bigr)\times\Gamma_n(i)\Bigr)\cup
\Bigl(\bigl(A\cap\Gamma_n(i)\bigr)\times\Pi_n(i)\Bigr)\biggr)$ and then:
$$\rien\hspace{-20pt}\begin{array}{@{}l@{}}\displaystyle
P_{n+1}\bigl(\mu_n(A)\bigr)=\sum_{i\in I_n}\left(
\sum_{\omega\in A\cap\Pi_n(i)}\left(\sum_{\omega'\in\Gamma_n(i)}
\frac{P_n(\omega)P_n(\omega')}{P_n\bigl(\Gamma_n(i)\bigr)}\right)
+
\sum_{\omega\in A\cap\Gamma_n(i)}\left(\sum_{\omega'\in\Pi_n(i)}
\frac{P_n(\omega)P_n(\omega')}{P_n\bigl(\Pi_n(i)\bigr)}\right)
\right)
\vspace{4pt}\\\displaystyle
\rien\hspace{100pt}
=\sum_{i\in I_n}\left(
\sum_{\omega\in A\cap\Pi_n(i)}P_n(\omega)+
\sum_{\omega\in A\cap\Gamma_n(i)}P_n(\omega)
\right)=
\sum_{\omega\in A}P_n(\omega)=P_n(A)\;.
\end{array}$$
\item[$\Box\Box\Box$]\rien
\end{description}
\emph{Corollary.}
$P_n(\Omega_n)=1$\,.
\\[3pt]
Derived from $P_0(\Omega_0)=\pi(\top)=1$ which is obvious.
\\[4pt]
\emph{Corollary of the corollary.}
$P_n$ is indeed a probability in the classical meaning.
\\[3pt]
Additivity, coherence are obtained by construction.
Finiteness comes from the corollary.
\subsubsection{Lemma 2}
\label{prop:3}
\begin{enumerate}
\item $\displaystyle P_n\bigl(\Pi_n(i)\bigr)+P_n\bigl(\Gamma_n(i)\bigr)=\frac{P_n\bigl(\Pi_n(i)\bigr)}{P_n(b_n)}=\frac{P_n\bigl(\Gamma_n(i)\bigr)}{P_n(\sim b_n)}$\,,
for any $i\in I_n$\,, 
\item
$P_{n+1}\bigl(\mu_n(b_n)\cap A\bigr)=P_{n+1}(\mu_n(b_n))P_{n+1}\Bigl(f_{n+1}\bigl(A,\mu_n(b_n)\bigr)\Bigr)$\,, for any $A\in \mathbf{B}_{n+1}$\,,
\item
$P_{n+1}\bigl(\sim \mu_n(b_n)\cap A\bigr)=P_{n+1}(\sim \mu_n(b_n))P_{n+1}\Bigl(f_{n+1}\bigl(A,\sim \mu_n(b_n)\bigr)\Bigr)$\,, for any $A\in \mathbf{B}_{n+1}$\,.
\end{enumerate}
These propositions are proved recursively.
\begin{description}
\item[Proof of 1.]
Obvious in case 1; the difficulty arises for case 0.
\\[3pt]
Assume now case 0, and let $(\omega,\omega')\in I_n$\,, \emph{i.e.} $\omega\in\mu_\nu(b_\nu)$ and $\omega'\in\sim\mu_\nu(b_\nu)$.
\\
Since $\mu_\nu(b_\nu)\cap\omega=\omega$, the recursion hypothesis over 2 yields:
$$\begin{array}{@{}l@{}}\displaystyle
P_{\nu+1}\Bigl(f_{\nu+1}\bigl(\omega',\sim \mu_\nu(b_\nu)\bigr)\cap\omega\Bigr)
\hspace{200pt}\rien
\gonextline\displaystyle
=P_{\nu+1}(\mu_\nu(b_\nu))P_{\nu+1}\biggl(f_{\nu+1}\Bigl(f_{\nu+1}\bigl(\omega',\sim \mu_\nu(b_\nu)\bigr)\cap\omega,\mu_\nu(b_\nu)\Bigr)\biggr)\,.
\end{array}$$
Finally $P_n\bigl(\Pi_n(i)\bigr)=P_n(b_n)P_{\nu+1}\Bigl(f_{\nu+1}\bigl(\omega',\sim \mu_\nu(b_\nu)\bigr)\cap f_{\nu+1}\bigl(\omega,\mu_\nu(b_\nu)\bigr)\Bigr)$\,.\\
Similarly, $P_n\bigl(\Gamma_n(i)\bigr)=P_n(\sim b_n)P_{\nu+1}\Bigl(f_{\nu+1}\bigl(\omega',\sim \mu_\nu(b_\nu)\bigr)\cap f_{\nu+1}\bigl(\omega,\mu_\nu(b_\nu)\bigr)\Bigr)$\,.\\
Then $\frac{P_n\bigl(\Pi_n(i)\bigr)}{P_n(b_n)}=\frac{P_n\bigl(\Gamma_n(i)\bigr)}{P_n(\sim b_n)}$ and the result is deduced from $P_n(b_n)+P_n(\sim b_n)=1$.
\item[Proof of 2.]
Since $P_{n+1}\bigl(T(\omega)\bigr)=P_{n+1}(\omega)\frac{P_n\bigl(\Gamma_n(i)\bigr)}{P_n\bigl(\Pi_n(i)\bigr)}$ for $\displaystyle\omega\in \bigcup_{i\in I_n}\bigl(\Pi_n(i)\times\Gamma_n(i)\bigr)$\,,
it comes:
$$\begin{array}{@{}l@{}}
\displaystyle P_{n+1}\Bigl(f_{n+1}\bigl(A,\mu_n(b_n)\bigr)\Bigr)=\sum_{i\in I_n}\biggl(\sum_{\omega\in A\cap(\Pi_n(i)\times\Gamma_n(i))}P_{n+1}(\omega)\frac{P_n\bigl(\Pi_n(i)\bigr)+P_n\bigl(\Gamma_n(i)\bigr)}{P_n\bigl(\Pi_n(i)\bigr)}\biggr)
\vspace{4pt}\\\displaystyle
\rien\hspace{50pt}
=\sum_{i\in I_n}\sum_{\omega\in A\cap(\Pi_n(i)\times\Gamma_n(i))}\frac{P_{n+1}(\omega)}{P_n(b_n)}
=\sum_{i\in I_n}\frac{P_{n+1}\Bigl(A\cap\bigl(\Pi_n(i)\times\Gamma_n(i)\bigr)\Bigr)}{P_n(b_n)}
\vspace{8pt}\\\displaystyle
\rien\hspace{50pt}
=\frac{P_{n+1}\bigl(\mu_n(b_n)\cap A\bigr)}{P_{n+1}(\mu_n(b_n))}\;,
\end{array}$$
by using 1.
\item[Proof of 3.] Similar to 2.
\item[$\Box\Box\Box$]\rien
\end{description}
\subsubsection{Conclusion.}
\label{AppC:Conclude}
Lemma~1 make possible the definition of $P_\infty:\mathbf{B}[\Theta]\rightarrow\Rset^+$ by:
\begin{equation}\label{dbl:ext:prob:eq:0:1}
P_\infty(\nu_n(A))=P_n(A)\,,\mbox{ for any }n\in\Nset\mbox{ and }A\in \mathbf{B}_n\;.
\end{equation}
Beside, $P_\infty$ is entirely defined then, since $\mathbf{B}[\Theta]=\bigcup_{n\in\Nset}\nu_n(\mathbf{B}_n)$\,.
\\[5pt]
By inheritance from the probabilities $P_n$ and lemma~2, $P_\infty$ verifies for any $A,B\in \mathbf{B}[\Theta]$\,:
\begin{equation}\label{dbl:ext:prob:prop:eq:1}\left.\begin{array}{@{}l@{}}
P_{\infty}(\emptyset)=0\mbox{ and }P_{\infty}(\Omega)=1\;,
\vspace{4pt}\\\displaystyle
P_{\infty}(A)+P_{\infty}(B)=P_{\infty}(A\cap B)+P_{\infty}(A\cup B)\,,
\quad\mbox{for any }A,B\in \mathbf{B}[\Theta]\;
\vspace{4pt}\\\displaystyle
P_{\infty}(A\cap B)=P_{\infty}(A)P_{\infty}\bigl(f_{\infty}(B,A)\bigr)\,,\quad\mbox{for any }A,B\in \mathbf{B}[\Theta]\;.
\end{array}\qquad\right]\end{equation}
Now, let us define the atomic assignment $h:\Theta\rightarrow\mathbf{B}[\Theta]$ by:
$$
h(\theta)=\nu_0(\xi_\theta)\,,\mbox{ where }
\xi_\theta=\bigl\{(\delta_\tau)_{\tau\in\Theta}\in\Omega_0\,\big/\,\delta_\theta=1\bigr\}\;.
$$
It is noticed, that its extention over $\mathcal{L}$, denoted $\overline{h}$, verifies:
\begin{equation}\label{dbl:ext:prob:eq:1}
\overline{h}\left(\bigwedge_{\tau\in\Theta}\tau_\omega\right)=\nu_0(\omega)\mbox{ for any }\omega\in\Omega_0\,,
\end{equation}
where $\tau_\omega$ is defined in~(\ref{dbl:apxF:deftauom:1}).
\\[5pt]
Then, define $\overline{\pi}$ for any $\phi\in\mathcal{L}$ by:
\begin{equation}\label{dbl:ext:prob:def:eq:1}
\overline{\pi}(\phi)=
P_\infty\bigl(\overline{h}(\phi)\bigr)\,.
\end{equation}
By property~(\ref{dbl:ext:prob:eq:1}) and the definition of $P_0$, it comes:
$$
\overline{\pi}(\phi)=\pi(\phi)\quad\mbox{for any }\phi\in\mathcal{L}_C\,.
$$
Now, from property~(\ref{dbl:ext:prob:prop:eq:1}), it comes that $\overline{\pi}$ is a multiplicative probability over $\mathcal{L}$.
\paragraph{Rational structure of $\overline{\pi}$\,.}
Let $\Sigma=\left\{\left.\bigwedge_{\theta\in\Theta}\epsilon_\theta\;\right/\;\epsilon\in\prod_{\theta\in\Theta}\{\theta,\neg\theta\}\right\}$\,.
For any $\phi\in\mathcal{L}$\,, there is a rational function $R_\phi: \Rset^{\Sigma}\rightarrow\Rset$ such that $\overline{\pi}(\phi)=R_\phi\bigl(\pi(\sigma)|_{\sigma\in\Sigma}\bigr)$\,.
\\[5pt]
The proof is obvious from the construction.
\end{document}